\documentclass[]{emulateapj}
\usepackage{natbib}
\usepackage{hyperref}
\usepackage{enumitem}
\hypersetup{
    colorlinks=true,
    linkcolor=blue,
    citecolor=blue,
    filecolor=blue,
    urlcolor=blue
}
 \providecommand{\adsurl}[1]{\href{#1}{ADS}}
\usepackage{amsmath,amssymb}
\usepackage{mathptmx}
\usepackage{mathtools}
\DeclareMathAlphabet{\pazocal}{OMS}{zplm}{m}{n}
\DeclareSymbolFont{matha}{OML}{txmi}{m}{it}% txfonts
\DeclareMathSymbol{\varv}{\mathord}{matha}{118}

\usepackage{graphicx}

\def\msun{\ifmmode {\rm M_{\odot}}\else $\rm M_{\odot}$\fi}
\def\mpc{\ifmmode {\rm M_{\odot} \ pc^{-2}} \else $\rm M_{\odot} \ pc^{-2}$ \fi}
\def\iuv{\ifmmode {I_{\rm UV}}\else $I_{\rm UV}$ \fi}

\def\sg{\ifmmode \sigma_g \else $\sigma_g$ \fi}
\def\ms{\ifmmode \mathcal{M}_s \else $\mathcal{M}_s$ \fi}
\def\fl{\ifmmode y_{\rm dec} \else $y_{\rm dec}$ \fi}
\def\lc{\ifmmode L_{\rm cloud} \else $L_{\rm cloud}$ \fi}
\def\ar{\ifmmode {\rm ArH^+} \else ArH$^+$ \fi}
\def\oh{\ifmmode {\rm OH^+} \else OH$^+$ \fi}
\def\ho{\ifmmode {\rm H_2O^+} \else H$_2$O$^+$ \fi}

\defcitealias{Bialy2017}{BBS}
\defcitealias{Neufeld2016}{NW}
\defcitealias{Neufeld2017b}{NW17}

\begin{document}

\title{Chemical Abundances in a Turbulent Medium -- 
H$_2$, OH$^+$, H$_2$O$^+$, A\MakeLowercase{r}H$^+$}
\author{Shmuel Bialy\altaffilmark{1}$^\star$, David Neufeld\altaffilmark{2}, Mark Wolfire\altaffilmark{3}, Amiel Sternberg\altaffilmark{4} \altaffilmark{5} \altaffilmark{6}, and Blakesley Burkhart\altaffilmark{5} \altaffilmark{7}}
\altaffiltext{1}{Harvard-Smithsonian Center for Astrophysics, 60 Garden street, Cambridge, MA, USA}
\altaffiltext{2}{Department of Physics and Astronomy, Johns Hopkins University, Baltimore, MD 20910, USA}
\altaffiltext{3}{University of Maryland, Astronomy Department, College Park, MD 20742, USA}
\altaffiltext{4}{School of Physics \& Astronomy, Tel Aviv University, Ramat Aviv 69978, Israel}
\altaffiltext{5}{Center for Computational Astrophysics, Flatiron Institute, 162 5th Ave., New York, NY, 10010, USA}
\altaffiltext{6}{Max-Planck-Institut f\"ur extraterrestrische Physik (MPE), Giessenbachstr., 85748 Garching, FRG, Germany}
\altaffiltext{7}{Department of Physics and Astronomy, Rutgers University,  136 Frelinghuysen Rd, Piscataway, NJ 08854, USA}
\email{$^\star$sbialy@cfa.harvard.edu}
\shorttitle{ArH$^+$, OH$^+$, H$_2$O$^+$ in a Turbulent Medium}
\shortauthors{Bialy et al.}
\slugcomment{The Astrophysical Journal. Submitted: Jul.~18, 2019.  Accepted: Sept.~25, 2019}

\begin{abstract}
Supersonic turbulence results in strong density fluctuations in the interstellar medium (ISM), which have a profound effect on the chemical structure. 
Particularly useful probes of the diffuse ISM are the ArH$^+$, OH$^+$, H$_2$O$^+$ molecular ions, 
which are highly sensitive to fluctuations in the density and the H$_2$ abundance. 
We use isothermal magnetohydrodynamic (MHD) simulations of various sonic Mach numbers, $\mathcal{M}_s$, and density decorrelation scales, $y_{\rm dec}$, to model the turbulent density field.
We post-process the simulations with chemical models and obtain the probability density functions (PDFs) for the H$_2$, ArH$^+$, OH$^+$ and H$_2$O$^+$ abundances.
We find that the PDF dispersions  increases 
with increasing $\mathcal{M}_s$ and $y_{\rm dec}$, as the magnitude of the density fluctuations increases, and as they become more coherent.
Turbulence also affects the median abundances:
when $\mathcal{M}_s$ and $y_{\rm dec}$ are high, low density regions with low H$_2$ abundance become prevalent, resulting in an enhancement of ArH$^+$ compared to OH$^+$ and H$_2$O$^+$.
We compare our models with {\it Herschel} observations. 
The large scatter in the observed abundances, as well as the high observed ArH$^+$/OH$^+$ and ArH$^+$/H$_2$O$^+$ ratios are 
naturally reproduced
 by our supersonic $(\mathcal{M}_s=4.5)$, large decorrelation scale $(y_{\rm dec}=0.8)$ model, supporting a scenario of a large-scale turbulence driving.
The abundances also depend on the UV intensity, CR
ionization rate and the cloud column density, and the observed scatter may be influenced by fluctuations in these parameters.
\end{abstract}


%Are the errors on the observations 1$\sigma$?
%If so, it is inconsistent with what we show for the theoretical 2D-PDFs, where we show 95 $\%$ contours ($=2\sigma$ for a Gaussian).

%Naming: Throughout the paper I use ``abundances" in different contexts, both to indicate $x(i) \equiv N(i)/n$, and 
%to indicate $N(i)/N({\rm H})$.
%How to distinguish the two?

%For more questions search the  latex text for XXX.

%Explain that the simulation is used to model different sighlines seperated with large distances, by probing randomly sightlines from the simluations.

\section{Introduction}
\label{sec: intro}

% {\color{magenta} David/Mark:
% General intro (molecules in the ISM, star-formation, etc.). 
% Motivations to study ArH+, OH+ H2O+ [references]}
% \newline\newline\newline\newline

A diverse collection of molecules and molecular ions has been detected in the interstellar medium (ISM).
The abundances of observed species and their ratios are often used to constrain cloud properties: the temperature, incident ultraviolet (UV) radiation flux, cosmic-ray ionization rate (CRIR), and the density of H nuclei, $n \equiv n({\rm H}) + 2n({\rm H_2})$.

Diffuse clouds in the ISM are commonly analyzed by chemical models for photodissociation-regions (PDRs; e.g., \citealt{Kaufman1999, Tielens1985a, Sternberg1995, Bell2006, Rollig2006, LePetit2006}).
\citet[][hereafter \citetalias{Neufeld2016}]{Neufeld2016} used constant-$n$  PDR models to study the abundances of ArH$^+$, OH$^+$, and H$_2$O$^+$, and compared their model predictions with observations.
These molecular-ions are sensitive to the H$_2$ abundance, which in turn depends on the 
 the UV intensity, the CRIR, the cloud visual extinction, and the gas density.
% \footnote{
% Hereafter $A_{\rm V}$ refers to the {\it total} $A_{\rm V}$ of the cloud, not to locations inside the cloud (as in \citetalias{Neufeld2016}).
% }.
\citetalias{Neufeld2016} showed that for all sightlines, there is a discrepancy between their theoretical models and the observations, 
as the models always under-predict the ArH$^+$ abundance relative to OH$^+$ and H$_2$O$^+$. 
% Any model that reproduces the observed OH$^+$ and H$_2$O$^+$ abundances, strongly under-predicts the ArH$^+$ abundance.
To explain this discrepancy, they were forced to invoke a two-cloud population model in which low $A_{\rm V}$ clouds were the source for the ArH$^+$, while higher $A_{\rm V}$ clouds produced the OH$^+$ and H$_2$O$^+$.
Furthermore, the observations show scatter in the column density ratios $N({\rm ArH^+})/N({\rm H})$, $N({\rm OH^+})/N({\rm H})$, and $N({\rm H_2O^+})/N({\rm H})$, for different lines-of-sight (LoS).
\citetalias{Neufeld2016} attributed the scatter to variations in the CRIR.

A common simplifying assumption in PDR models is that the gas density in the cloud is constant (or is smoothly varying with cloud depth).
However, this assumption may be wrong.
Observations suggest that molecular clouds as well as cold atomic clouds exhibit turbulent supersonic motions, as is evident from, e.g.: the observed superthermal
linewidths, the line-width size relation, the fractal structures of molecular clouds, and studies of the power spectrum/bispectrum \citep[e.g., ][]{Vazquez-Semadeni1994a, Stutzki1998, Sanchez2005, Heyer2009, Burkhart2009, Roman-Duval2010, Chepurnov2015}. %this review is very outdated and doesn't discuss more modern metrics of turbulence, e.g the paragraph I wrote -BB
%try these please 
%(see Elmegreen \& Elmegreen
% 1983; Vazequez-Semandeni 1994; Stutzki et al. 1998;
% Sanchez et al. 2005; Roman-Duval et al. 2010; Heyer et al. 2009; Goodman et al. 2011; Burkhart et al. 2009; Burkhart et al. 2013; Chepurnov et al. 2015; Pingel et al. 2017; Portilo et al. 2017).
% SB: hi Blakes, could you please specify which Goodman2011 , Burkhart2013 , Pingel2017 and Portilo2017 papers did you mean?
% also, could you please verify that the above papers I cited are the ones you meant?
The turbulence, when supersonic, produces strong density fluctuations in the gas, which alter the chemical structure. 

\citet[][hereafter \citetalias{Bialy2017}]{Bialy2017} have shown that in a turbulent medium, density fluctuations 
perturb the H and H$_2$ abundances in the gas,
resulting in large dispersions in the atomic columns, $N({\rm H})$,
for different LoS. 
The variance in the $N({\rm H})$ distribution was related to the governing turbulent parameters: the sonic Mach number, $\ms$, and the decorrelation scale of the density field, $L_{\rm dec}$, which is proportional to the turbulence driving scale, $L_{\rm drive}$.
% This opens up the avenue to using observations of chemical abundances in and around clouds, to constrain the nature of turbulence.

In this paper we extend upon the \citetalias{Neufeld2016} and \citetalias{Bialy2017}  analyses. 
We study how turbulence affects the abundances of H$_2$, ArH$^+$, OH$^+$, and H$_2$O$^+$, and compare our models with observations.
We find that our turbulent-cloud models may explain the observed scatter in the observed abundances, as well as the high ArH$^+$/OH$^{+}$ and ArH$^+$/H$_2$O$^{+}$ ratios.
We demonstrate how the observations may be used to constrain the turbulent parameters, $\ms$ and $L_{\rm drive}$.

% We combine chemical PDR models with magnetohydrodynamic (MHD) simulations and model the abundances as functions of turbulent parameters: the sonic Mach number, and the density decorrelation scale (i.e., the density fluctuation size-scale). 
% We discuss how the density fluctuations alter the chemical abundances (compared to uniform-density models).
% We find that the turbulent-cloud model alleviates the tension between the ArH$^+$ and OH$^+$, H$_2$O$^+$ observations.
% Furthermore, we show that the turbulent density fluctuations naturally produce a scatter in the species abundances among sightlines, thus strongly reducing the required scatter in the cosmic-ray (CR) ionization rate.

% However, an important simplifying assumption invoked by the \citetalias{Neufeld2016} models is that the gas has a uniform density, $n$.
% While this is a common assumption in the literature, 

The paper is organized as follows.
In  \S \ref{sec: chemistry}  we review the chemical network and discuss the governing physical parameters.
% relevant to H, H$_2$, ArH$^+$, OH$^+$ and H$_2$O$^+$.
In \S \ref{sec: model} we discuss the model ingredients, including our MHD simulations and the chemical PDR models. 
In \S \ref{sec: results} we present our results for the abundance distributions in turbulent clouds.
In \S \ref{sec: obs} we compare our models
with observations.
We discuss the limitations of the model and our conclusions in \S \ref{sec: Discussion} and \S \ref{sec: conclusions}.

\section{Chemistry}
\label{sec: chemistry}

% We calculate the chemical abundances and column densities of H, H$_2$, ArH$^+$, OH$^+$, and H$_2$O$^+$ for turbulent, highly inhomogeneous gas, as modeled by magnetohydrodynamical (MHD) simulations. 
% 
The formation and destruction pathways include two-body reactions, surface chemistry on dust-grains, and cosmic-ray and UV photoreactions.
% We assume chemical steady-state.
We define the abundance of species $i$, $x({\rm i}) \equiv n({\rm i})/n$, where $n(i)$ is the density of species $i$, and $n$ is the density of hydrogen nuclei.
% \footnote{Subscript names refer to elements, whereas names in parentheses refer to species.}.

\subsection{H and H$_2$}
\label{subsub: H H2}

The H$_2$ is formed on the surfaces of dust grains.
The removal of H$_2$ is through photodissociation by Lyman-Werner (LW) radiation (11.2-13.6 eV),
and by cosmic-ray (CR) ionization.
The H and H$_2$ steady-state abundances
obey
\begin{equation}
\label{eq: H_H2 steady state}
\frac{x({\rm H_2})}{x({\rm H})} =  \frac{Rn}{D_0 f_{\rm att}+\zeta_{\rm tot}}  \ ,
\end{equation}
% \begin{equation}
% \label{eq: H conservation}
% \ ,
% \end{equation}
% In Eqs.~(\ref{eq: H_H2 steady state}-\ref{eq: H conservation}), $R$ is the H$_2$ formation rate coefficient, $D_0$ (s$^{-1}$) is the photodissociation rate by LW photons in free-space, $f_{\rm att} \leq 1$ is an attenuation factor that accounts for H$_2$ self-shielding and dust absorption, and $\zeta_{\rm tot}$ is the total destruction rate of H$_2$ arising from cosmic-ray (CR) ionizations, including secondary ionizations and chemical reactions with molecular ions which are produced by CR ionization.
and by element conservation, $x({\rm H}) + 2x({\rm H_2}) \simeq 1$.
Here, $R= 3 \times 10^{-17}$ cm$^3$ s$^{-1}$
% $R$
is the
 H$_2$ formation rate coefficient for 
 formation on dust-grains
(the value is for a 100 K gas),
and $D_0=5.8 \times 10^{-11} I_{\rm UV}$ s$^{-1}$ 
\citep[][]{Sternberg2014}, 
% $D_0$
where
$I_{\rm UV}$ is the interstellar radiation field in units of the value given by \citet{Draine1978}.
The attenuation factor, $f_{\rm att}$, accounts for the absorption of LW photons by dust and in H$_2$ lines.
For normal incidence radiation, penetrating a slab on two sides 
% propagating parallel to the $z$ coordinate, the local photodissociation rate 
%  at a point inside the box is
\begin{equation}
\label{eq: diss rate}
f_{\rm att} = \frac{1}{2}f_{\rm sh}(N^l({\rm H_2})) \mathrm{e}^{-\sigma_g N^l}  + \frac{1}{2}f_{\rm sh}(N^r({\rm H_2})) \mathrm{e}^{-\sigma_g N^{r}} \ .
\end{equation}
Here $\sigma_g =1.9 \times 10^{-21}$ cm$^2$  is the dust absorption cross section per hydrogen nucleus \citep{Sternberg2014}, and
$f_{\rm sh}$ is the H$_2$ self-shielding function 
\citep[Eq.~37 in][]{Draine1996} 
for which we adopt $b=2$ km/s (see \S \ref{sub: H$_2$ self-shielding} and appendix \ref{appendix doppler}).
$N$ and $N({\rm H_2})$ are the column densities of hydrogen nuclei and H$_2$, from cloud edge to the point of interest, and the superscripts $l$ and $r$ denote integration from the left and right cloud edges, respectively.
% The dust absorption and H$_2$ self-shielding depend on
% the accumulated columns of H$_2$ and H, from cloud edges to the point of interest,
% For sufficiently large column, photodissociation becomes negligible and H$_2$ removal is dominated by cosmic-rays (CRs).
% and $f_{\rm att}$ are the free-space photodissociation rate and the attenuation factor which accounts for H$_2$ self-shielding and dust absorption, and

In Eq.~(\ref{eq: H_H2 steady state}), $\zeta_{\rm tot} \approx 2.3 \times 1.9 \times  \zeta$ is the total CR removal rate of H$_2$ and where $\zeta$ is the primary CRIR of H.
The factor $2.3$ translates from primary CR ionization per H to total (primary+secondary) ionization per H$_2$ 
\citep{Glassgold1974}, and the additional factor $1.9$ accounts for H$_2$ destruction by chemical reactions with molecular ions which are produced by CR ionization (i.e., the H$_2$ abstraction reactions below)\footnote{
%The factors of 2.3 and 1.9 
%depend on the local 
%abundances and vary within the cloud. 
We have fixed the 2.3 and 1.9  factors at values appropriate for cloud interiors.
The value 1.9 is found by taking an average value 
%at cloud center %
from our modeling results accounting for the H$_2$ destruction from all chemical pathways.
Towards cloud edge, these factors vary, however
in these regions H$_2$ destruction is in any case dominated by UV photodissociation, not by CR processes.
}.
We define $\zeta_{-16} \equiv \zeta/(10^{-16} {\rm s^{-1}})$
% For a \citet{Draine1978} radiation spectrum shape $D_0=5.8 \times 10^{-11} I_{\rm UV}$ where $I_{\rm UV}$ is the UV intensity normalized to the \citep{Draine1978} UV field.

% Removal of H$_2$ by CR ionization becomes important at large cloud depths where the UV radiation does not efficiently penetrate.
% Each ionization leads to the formation of energetic electrons which lead to further secondary ionizations of  H$_2$.
% The destruction of H$_2$ is further enhanced by reactions with molecular ions which are also produced by CR ionization.
% Accounting for all these processes, the total removal rate of H$_2$ molecules by CR may be written as
% \begin{equation}
% \label{eq: zeta_tot}
% \zeta_{\rm tot} = \phi_1 \phi_2 \zeta_{\rm p, H} \ ,
% \end{equation}
% where $ \zeta_{\rm p, H}$ is the primary CR ionization rate of atomic hydrogen.
% The factor $\phi_1=2.3$ is the total (primary+secondary) ionizations per each primary ionization event.
% $\phi_2=xxx$ converts from the ionization rate per H to ionization per H$_2$ molecule. The factor $\phi_3=1.9$ accounts for additional destruction by the heavy element chemistry.
% %shmuel: We adopt phi1*phi2=2.3. What are the values seperately. I always though phi1~1.6 and phi2~2, so that I expect phi1*phi2=3.2. Why is it only 2.3?

\begin{figure}
	\centering
	\includegraphics[width=0.5\textwidth]{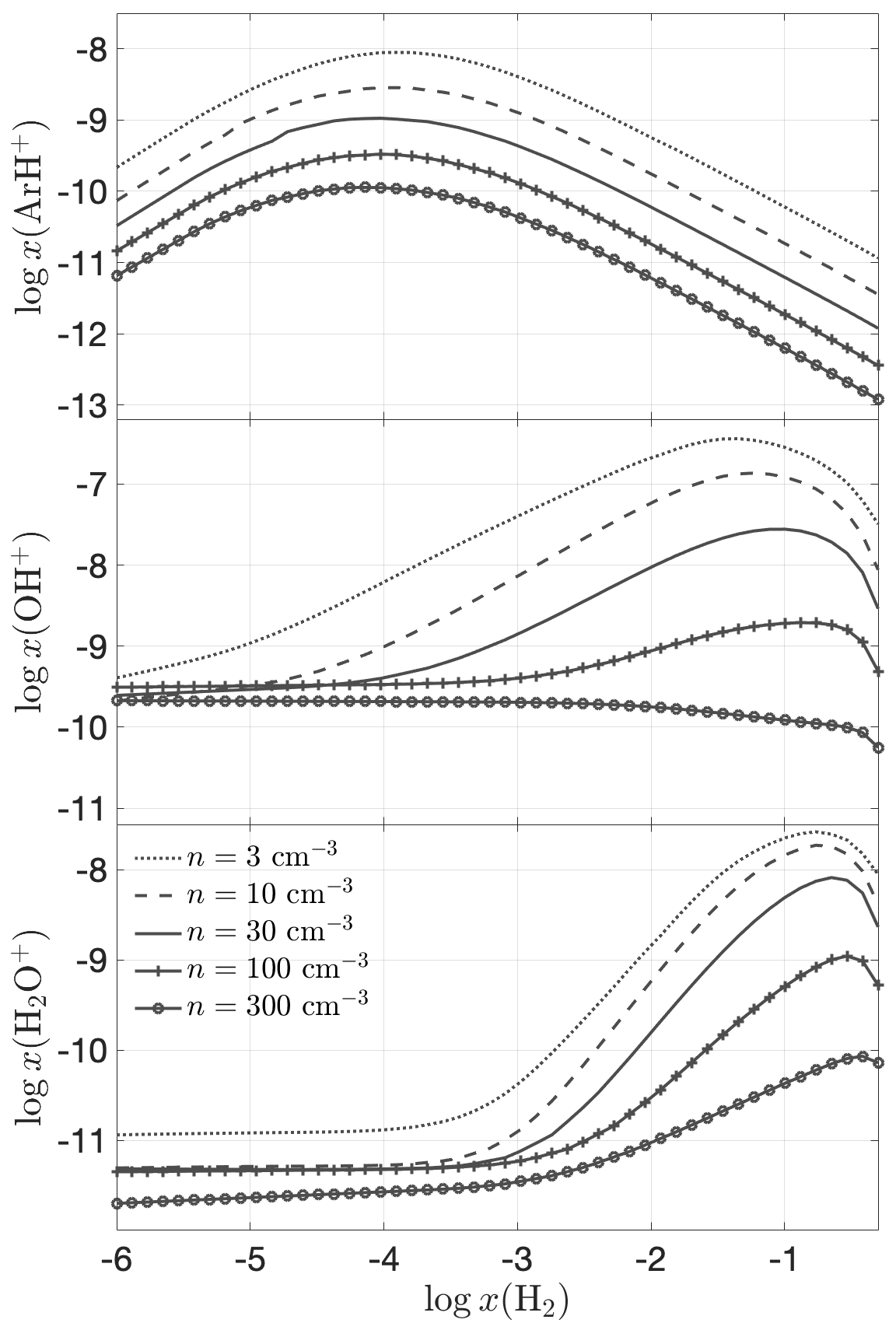} 
	\caption{
	The ArH$^+$, OH$^+$, and H$_2$O$^+$ abundances ($x(i) \equiv n(i)/n$) as functions of the H$_2$ abundance, for $I_{\rm UV}=1$, $\zeta_{-16}=4$, and various (constant) densities $n$.
		}
    \label{fig: x_vs_x2}
\end{figure}

\subsection{ArH$^+$, OH$^+$, and H$_2$O$^+$}
\label{subsub: ArHp et al}

Here we discuss the basic formation and destruction chemistry
for ArH$^+$, OH$^+$, and ${\rm H_2O^+}$ (see also
\citealt{Schilke2014} and \citetalias{Neufeld2016}).
The production of ArH$^+$ is initiated by CR ionization followed by hydrogen abstraction 
\begin{align}
\label{eq: Ar ionization}
   {\rm Ar} + {\rm cr} &\rightarrow {\rm Ar^+} + {\rm e} \ , \\
   \label{eq: Ar+ abstraction}
{\rm Ar^+} + {\rm H_2} &\rightarrow {\rm ArH^+} + {\rm H} \ .
\end{align}
The formation sequence is moderated by the recombination of Ar$^+$ with polycyclic aromatic hydrocarbons (PAH), and PAH$^-$ \citep[dielectric recombination is subdominant,][]{Arnold2015}.
Dissociative recombination of ArH+ is unusually slow
(\citetalias{Neufeld2016}), and its destruction is dominated by
proton transfer
with H$_2$
\begin{equation}
{\rm ArH^+} + {\rm H_2} \rightarrow {\rm Ar} + {\rm H_3^+} \ . 
\end{equation}
If the H$_2$ abundance is low, $x({\rm H_2})\lesssim 10^{-4}$, the removal of ${\rm ArH^+}$ proceeds mainly through charge transfer with O and C.

\begin{figure*}[t]
	\centering
	\includegraphics[width=1\textwidth]{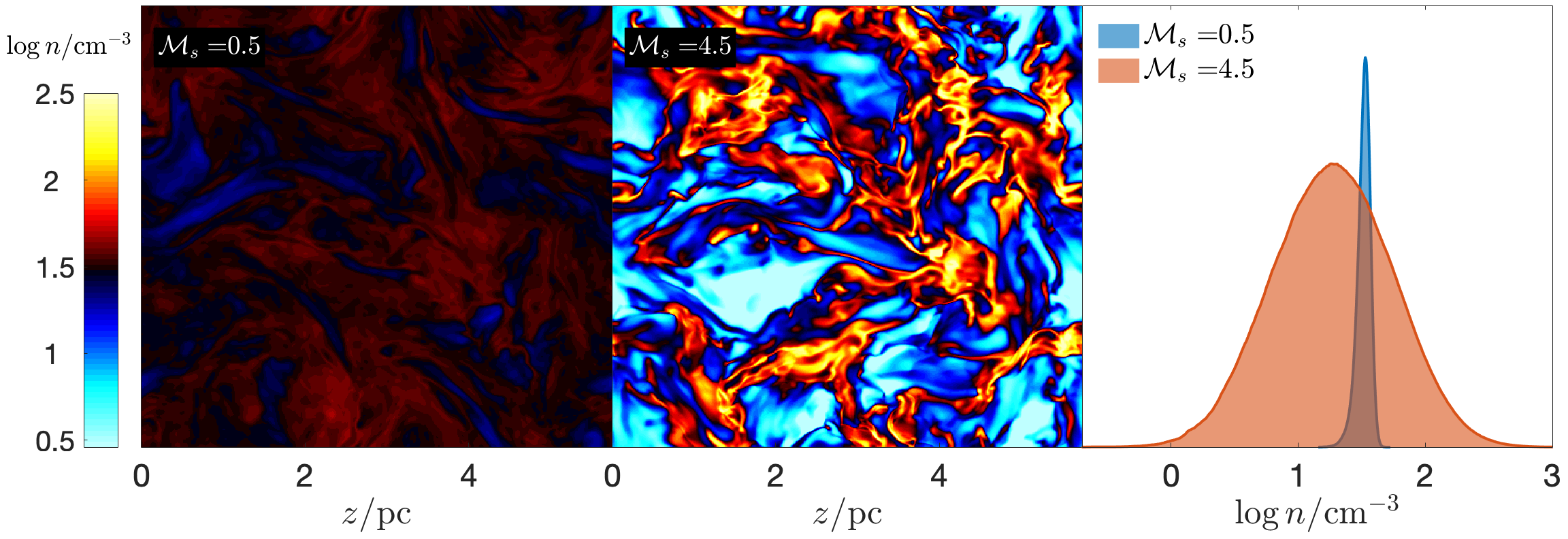} 
	\caption{
Density cuts through the $\ms=0.5$ and $4.5$ simulations (with $\fl=0.08$), and the corresponding density PDFs.
		}
\label{fig: sim_box}
\end{figure*}

The production of ${\rm OH^+}$ 
is initiated by CR ionization of H,  followed
by charge transfer and H abstraction,
\begin{align}
  \label{eq: H ionization}
 {\rm H} + {\rm cr} &\rightarrow {\rm H^+} + {\rm e} \ ,     \\
  \label{eq: H+ charge transfer}
 {\rm H^+} + {\rm O} &\rightarrow {\rm O^+} + {\rm H} \ ,  \\
 \label{eq: O+ abst to OH+}
 {\rm O^+} + {\rm H_2} &\rightarrow {\rm OH^+} + {\rm H} \ .
\end{align}
If the H$_2$ abundance is sufficiently high, OH$^+$ is formed through the sequence
\begin{align}
  \label{eq: H2 ionization}
{\rm H_2}+{\rm cr} &\rightarrow {\rm H_2^+} + {\rm e} \ , \\
  \label{eq: H3+ formation}
{\rm H_2^+}+{\rm H_2}&\rightarrow {\rm H_3^+} + {\rm H} \ , \\
  \label{eq: H3+ and O to OH+}
{\rm H_3^+}+{\rm O}&\rightarrow {\rm OH^+} + {\rm H_2} \ .
\end{align}
Further abstraction reactions with H$_2$ destroy OH$^+$ and lead to the formation of H$_2$O$^+$ and H$_3$O$^+$
\begin{align}
\label{eq: OH+ abst to H2O+}
{\rm OH^+} + {\rm H_2} &\rightarrow {\rm H_2O^+} + {\rm H} \ , \\
\label{eq: H2O+ abst to H3O+}
{\rm H_2O^+} + {\rm H_2} &\rightarrow {\rm H_3O^+} + {\rm H} \ .
\end{align}
OH+ and H2O+ are also destroyed  by dissociative recombination and photodissociation (see Fig.~2 in
\citealt{Bialy2015a}).

Because of the abstraction reactions with H$_2$, the OH$^+$ and H$_2$O$^+$ abundances are very sensitive to $x({\rm H_2})$, down to very low H$_2$ abundances, $x({\rm H_2}) \sim 10^{-3} - 10^{-4}$. 
At still
lower ${\rm H_2}$ abundances, OH$^+$ and H$_2$O$^+$ are formed via solid-state chemistry on dust grains
\citep{Hollenbach2012,Sonnentrucker2015}, and their abundances are independent of $x({\rm H_2})$. 
In this
regime, atomic hydrogen on grains reacts with atomic oxygen to form OH. The OH
can be desorbed from the grain surface or react with another hydrogen to form 
${\rm H_2O}$. 
Upon desorption, OH and ${\rm H_2O}$ undergo a charge 
transfer   
with ${\rm H^+}$ forming ${\rm OH^+}$ and ${\rm H_2O^+}$.
% \begin{align}
% {\rm OH}+{\rm H^+}&\rightarrow {\rm OH^+} + {\rm H} \ , \\ \nonumber
% {\rm H_2O}+{\rm H^+}&\rightarrow {\rm H_2O^+} + {\rm H} \ .
% \end{align}
% The result 
% of grain surface
% reactions is to produce a floor in the ${\rm OH^+}$ and ${\rm H_2O^+}$ 
% abundances which
% do not further diminish with decreasing $x({\rm H_2})$.  

In Fig.~\ref{fig: x_vs_x2} we show the ${\rm ArH^+}$, ${\rm OH^+}$ and ${\rm H_2O^+}$ abundances as functions of $x({\rm H_2})$, as computed by our PDR models (see \S \ref{sub: PDR} below)
for $I_{\rm UV}=1$, $\zeta_{-16}= 4$, and for various densities, $n$.
For a fixed $x({\rm H_2})$ the abundances of all the molecular ions increase with decreasing $n$. 
The formation rate is proportional to $\zeta$, while destruction is proportional to $n$, so all three molecular ions have abundances that increase with $\zeta/n$.

At a constant $n$, the $x({\rm ArH^+})$ is a non-monotonic function of $x({\rm H_2})$: it increases with $x({\rm H_2})$ when $x({\rm H_2})$ is very small, peaks at $x({\rm H_2}) \sim 10^{-4}$, and then decreases with $x({\rm H_2})$.
This behavior may be understood as follows. 
When $x({\rm H_2})$ is very small, the abstraction reaction (\ref{eq: Ar+ abstraction}) is slow, and is heavily interrupted by 
Ar$^+$ recombinations with PAHs.
As a result, the formation rate is $\propto \zeta$ multiplied by a branching ratio $\propto x({\rm H_2})$.
Destruction, on the other hand, is dominated by charge transfer with O and C, and is independent of $x({\rm H_2})$. Thus the abundance scales with $x({\rm H_2})$.
At high $x({\rm H_2})$, the abstraction reaction proceeds rapidly such that every Ar ionization leads to the formation of ArH$^+$, and the formation is independent of $x({\rm H_2})$.
However, removal is now dominated by interactions with H$_2$.
Thus the overall abundance scales inversely with $x({\rm H_2})$. 
Similar considerations explain the non-monotonic behavior of the OH$^+$ and H$_2$O$^+$ abundances.

\section{Model Ingredients}
\label{sec: model}

\subsection{Chemical PDR Models}
\label{sub: PDR}

We ran a large series of chemical PDR models similar to those described by \citet{Wolfire2010} and \citet[][with updates
as described in \citetalias{Neufeld2016}]{Hollenbach2012}.
Each model is characterized by (1) $I_{\rm UV}/n$, (2) $\zeta/n$, and (3) total visual extinction, $A_{\rm V}$, through the cloud (denoted $A_{\rm V}{\rm (tot)}$ by \citetalias{Neufeld2016}).
% We hereafter use the normalized parameters (1) $\pazocal{I} \equiv \frac{I_{\rm UV}}{n/({\rm 30 cm^{-3}})}$ and (2) $\pazocal{Z} \equiv \frac{\zeta/({\rm 4 \times 10^{-16} s^{-1}})}{n/({\rm 30 cm^{-3}})}$,
% normalized to our fiducial parameter values.
The models span the parameter space $I_{\rm UV}/n=20$ to $10^{-3}$ cm$^{3}$, $\zeta/n= 10^{-20}$ to $4 \times 10^{-16}$ cm$^3$ s$^{-1}$, and $A_{\rm V}=3\times 10^{-4}$ to $8$ mag.
For any given set of parameters, the model computes the temperature and molecular abundances (assuming steady-state) as functions of cloud depth, hereafter the $z$ coordinate.

Generally the fractional densities inside the cloud, $x(i)$, depend on the four parameters, $I_{\rm UV}/n$, $\zeta/n$, the total cloud visual extinction $A_{\rm V}$, and cloud depth $z$.
However, for ArH$^+$, OH$^+$, and H$_2$O$^+$, the dependence on $z$ and $A_{\rm V}$ is fully captured by the variation of the H$_2$ fraction, $x({\rm H_2})$.
To this end, 
we re-express the molecular-ion abundances as functions of $x({\rm H_2})$, reducing the rank of the parameter-space by 1  ($\{z,A_{\rm V} \} \rightarrow x({\rm H_2})$). 
We construct lookup tables for $x(i)$ ($i={\rm ArH^+, OH^+, H_2O^+}$) as functions of $I_{\rm UV}/n$, $\zeta/n$ and $x({\rm H_2})$.

% , and an analytic fit for ArH$^+$
% \begin{equation}
% \label{eq: xArH+}
%     x({\rm ArH^+}) = 
%      \frac{1.6  \times 10^{-9} \pazocal{Z}  }{\left(1+ 0.22\ \frac{1 + 0.28 \pazocal{Z}}{x_{-4}}\right) \left(1 + 0.25 x_{-4} \right)}  .
% \end{equation}
% Here $\pazocal{Z} \equiv (\zeta/4 \times 10^{-16} {\rm s^{-1}})/(n/30 {\rm cm^{-3}})$ and $x_{-4} \equiv x({\rm H_2})/10^{-4}$. \footnote{This fit is accurate to better than 8\% for 
% $x_{-4} > 1.5$, 27\% for 
% $x_{-4} > 50$, and a factor 2 for all 
% $x_{-4}$. . }

% In Eq.~(\ref{eq: xArH+}) and in our lookup tables we have 
% transalted the dependence on $A_{\rm V, tot}$ and on cloud depth through the calculated H$_2$ abundance in the models, $x({\rm H_2})$.
% This parameterization reflects the strong dependence of the ArH$^+$, OH$^+$, and ${\rm H_2O^+}$ abundances 
% on the H$_2$ content.

\subsection{MHD Simulations}
\label{sub: MHD}

% In non-dimensional form, these are
% \begin{align}
%  \frac{\partial x}{\partial \bar{t}} + \bar{\nabla} \cdot (x {\bf u}) &= 0, \\
%  \frac{\partial (x {\bf u})}{\partial \bar{t}} + \bar{\nabla} \cdot \left[ \rho {\bf u} {\bf u} + \left( \ms + 2\pazocal{M}_A^2  \right) {\bf I} - \pazocal{M}_A^2 {\bf b}{\bf b} \right] &= \bar{\bf f}, \\ 
%  \frac{\partial {\bf b}}{\partial \bar{t}} - \bar{\nabla} \times ({\bf u} \times{\bf b}) &= 0 \ ,  
% \end{align}
To model the density field in the turbulent medium, we solve the ideal MHD equations
\begin{align}
\frac{\partial \rho}{\partial t} + \nabla \cdot (\rho \pmb{\varv}) &= 0 \ , \\
 \frac{\partial \rho \pmb{\varv}}{\partial t} + \nabla \cdot \left[ \rho \pmb{\varv} \pmb{\varv} + \left( p + \frac{B^2}{8 \pi} \right) {\bf I} - \frac{1}{4 \pi}{\bf B}{\bf B} \right] &= {\bf f} \ ,  \\
 \frac{\partial {\bf B}}{\partial t} - \nabla \times (\pmb{\varv} \times{\bf B}) &= 0 \ .
\end{align}
Here $\rho$ is density,
% \footnote{In the remainder of the text we denote the normalized volume density as $x\equiv n/\langle n \rangle$, where $\langle n \rangle$ is the mean volume density.}
${\bf B}$ is magnetic field, $p$ is the gas pressure, ${\bf I}$ is the identity matrix and $\bf{f}$ is the specific force.
We use a third-order-accurate hybrid essentially nonoscillatory scheme on a 3D grid of 512$^3$  cells.
The $B$ field is initially aligned along the $y$ axis.
We assume 
 periodic boundary conditions, and an isothermal equation of state $p = c_s^2 \rho$.
By constrained transport we numerically construct $\nabla \cdot {\bf B} = 0$.
For the source term $\bf{f}$, we assume a random large-scale solenoidal driving at a wave number $k\approx 2.5$ (i.e.~1/2.5 the box size). 
For more details see \citet{Burkhart2009}.

Each simulation is defined by two parameters, the sonic and Alfv\'{e}nic Mach numbers, $\ms \equiv  \langle |\pmb{\varv}| \rangle/c_s$, $\mathcal{M}_A \equiv \langle |\pmb{\varv}| \rangle/ \varv_A $, where $\pmb{\varv}$ is the velocity, $c_s$  and $\varv_A$ are the isothermal sound speed and the Alfv\'en speed, and $\langle \cdot \rangle$ denotes averages over the entire simulation box.
We consider three simulations, $\ms =0.5$, 2.0 and 4.5, corresponding to subsonic, transonic, and supersonic medium. All simulations have $\mathcal{M}_A = 2$.
As discussed in \citetalias{Bialy2017}, $\mathcal{M}_s$ strongly affects the variance of the density field,
 and hence the H-H$_2$ structure.
 However, the H-H$_2$ structure is only weakly sensitive to the Alfv\'{e}nic number.
As the simulations do not include gravity and chemistry the density and size are scale-free 
\citep[see Appendix in][]{Hill2008}.
We adopt $\lc = 6.3$ pc for the simulation box length, and $\langle n \rangle = 30$ cm$^{-3}$, corresponding to a mean cloud column $\langle N \rangle \equiv \langle n \rangle \lc = 5.8 \times 10^{20}$ cm$^{-2}$ or $\langle A_{\rm V} \rangle=0.3$ mag.
% \footnote{$A_{\rm V}=0.53 \ {\rm mag} \ N/(10^{21} {\rm cm^{-2}})$}.

In Figure \ref{fig: sim_box} we show density slices, and density PDFs, for the $\ms=0.5$ and 4.5 simulations.
The density is nearly homogeneous in the subsonic simulation, while in the supersonic case shocks develop and lead to strong density fluctuations in the gas.
% In the $\ms=4.5$, low density gas, with $n < \langle n \rangle$ occupy most of the volume, 
% where the density may fall to values as low as $\sim 1$ cm$^{-3}$.
% Shock-compressed regions may reach densities as high as $\sim 1000$ cm$^{-3}$.
The $n$-PDF is close to a lognormal, 
with a dispersion that increases with the sonic Mach number,
as $\sigma_{\ln (n/\langle n \rangle)}^2 \simeq \ln [1+ (b \ms)^2]$, where
$b=1/3$ is the forcing parameter \citep{Federrath2008}.

% \begin{equation}
% \label{eq: sigx}
% \sigma_{\ln x}^2 \simeq \ln [1+ (b \ms)^2]  \ .
% \end{equation} 
% as $\sigma_{\ln \rho}^2 \simeq \ln [1+ (b \ms)^2]$, and where $\mu_{\ln \rho} = -\sigma_{\ln \rho}^2/2$.
% Here, $b$ parameterizes the driving force with $b=1/3$ for pure solenoidal driving, and $b=1$ for pure compressional driving. In our simulations, $b=1/3$ (cite XXX).
% Following the Gaussian distribution properties, $\mu_{\ln x} = -\sigma_{\ln x}^2/2$, and $\sigma_x = b \ms$, and  $x_{\rm median} =[1+(b\ms)^2]^{-1/2}$.
% The PDF shape and its dependence on $\ms$ are in agreement with previous studies (cite XXX).

The large-scale driving yields a density field that is correlated over large scales.
The decorrelation scale
 in the simulation is $\fl \equiv L_{\rm dec}/\lc = 0.08$, i.e., $\approx 40$ cells (\citetalias{Bialy2017}). 
$L_{\rm dec}$ is the characteristic scale over which
correlations in the density field decrease:
For averaging over length $L$, the density variance decreases with increasing $L$, and 
 $L_{\rm dec}$ is the characteristic scale for this decreasing trend
 (see \S 4.2 in \citetalias{Bialy2017}. cf.~\citealt[][]{VazquezSemadeni2001} for alternative definitions).
In supersonic gas, density fluctuations are still present on scales below $L_{\rm dec}$, all the way down to the sonic-length, $L_s$ (see footnote 10  below). 
As evident in Fig.~\ref{fig: sim_box}, $L_{\rm dec}$ is indeed a typical length-scale for the 
large-scale density modes
 in the simulations.
The value of $\fl$ has a strong influence on the resulting chemical structure (see \S \ref{subsub: fluc} below).
%  For example, as we discuss in \S \ref{subsub: fluc}, as $\fl$ increases the distribution of column abundances exhibit a larger scatter.
To study the effects of variations in $\fl$, we crop each of the simulations after 51 cells, producing thin slabs of thickness $1/10$ of the original boxes.
The relative decorrelation length in the slabs (along the short axes) is $\fl \simeq 0.8$. The slab thickness is still set to
$\lc = 6.3$ pc, as for the original simulation boxes.

% In total we explore six cases, 
% $\ms = 0.5, 2.0$ and 4.5, each of which with $\fl = 0.08$, and with 0.8.
\begin{figure*}[t]
	\centering
	\includegraphics[width=1\textwidth]{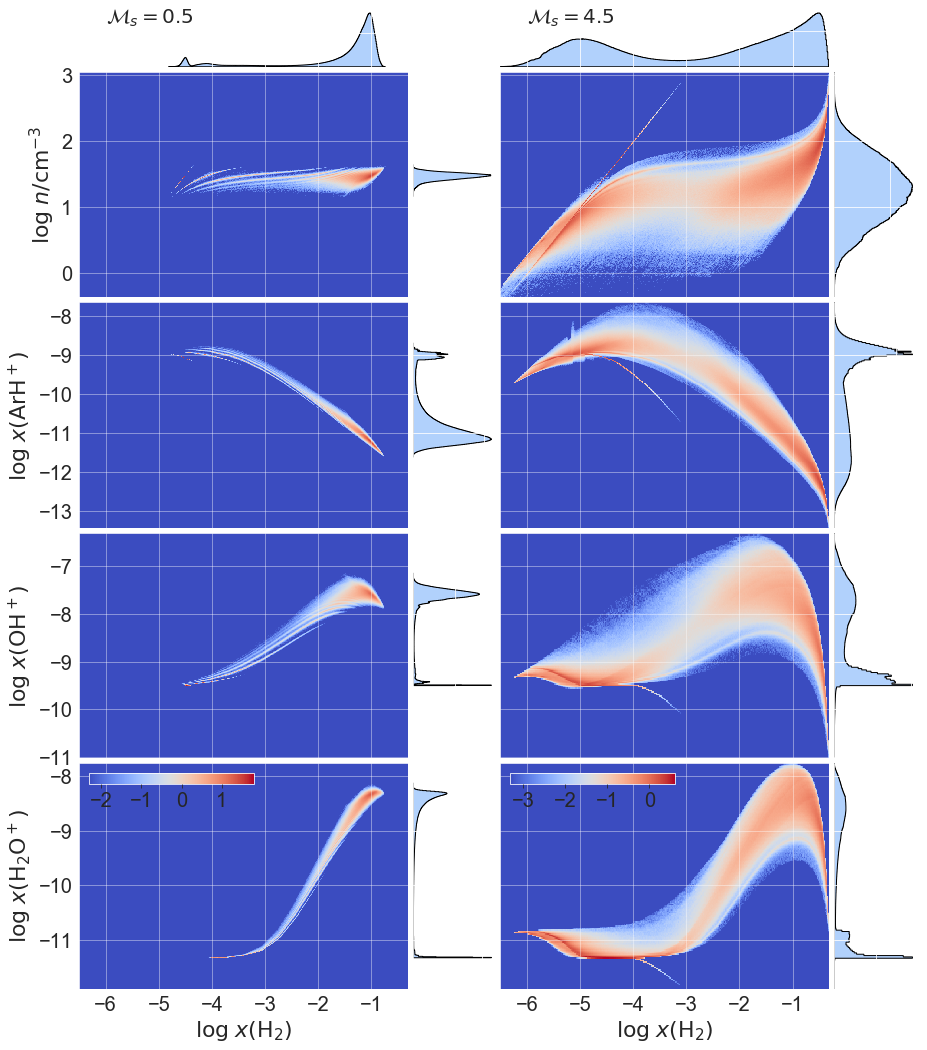}
	\caption{
	The joint-PDFs of $x({\rm H_2})$ versus $n$, $x({\rm ArH^+})$, $x({\rm OH^+})$ and $x({\rm H_2O^+})$, for the $\ms=0.5$ (left) and 4.5 (right) simulation boxes, with $\fl=0.08$, and $I_{\rm UV}=1$, $\zeta_{-16}=4$, $\langle n \rangle=30$ cm$^{-3}$, $\langle A_{\rm V} \rangle=0.3$.
	The colors corresponds to $\log$ of the PDF, as indicated (each of the colorbars applies to the entire column).
		The marginal PDFs (in linear scale) are also shown. 
	The stratification seen in the left panels is an artifact resulting from the finite size of simulation cells.
	In our chemical computations, we use a much finer (logarithmic) grid, and then we interporlate the results onto the simulation boxes. 
		}
    \label{fig: abundance PDFs}		
\end{figure*}

\subsection{Numerical Procedure}
\label{sub: numerical}

% At any point, the H and H$_2$ abundances depend on the column densities from cloud edges to the point of interest, and on the ratios $I_{\rm UV}/n$ and $\zeta/n$.
% In our models we adopt $I_{\rm UV}=1$
% %  \footnote{$1 I_{\rm UV}=1.7 G_0$ where $G_0$ is the UV intensity relative to the \citep{Habing1968} field.}.
% % In our models, we adopt $R = 3 \times 10^{-17}$ cm$^3$ s$^{-1}$, 
% % $I_{\rm UV}=1$,
% and $\zeta = 4 \times 10^{-16}$ s$^{-1}$.
% For $n$ we consider a non-uniform density gas
% with a mean density $\langle n \rangle = 30$ cm$^{-3}$, with strong density fluctuations ($\delta n \sim \langle n \rangle$; see XXX).
% Thus, the fluctuations in density result in fluctuations in $x({\rm H_2})/x({\rm H})$ compared to a uniform density gas \citep[][]{Bialy2017}.

As our focus is on the effects of turbulence, we fix the non-turbulent parameters
$I_{\rm UV}=1$, $\zeta_{-16}=4$, $\langle n \rangle  = 30$ cm$^{-3}$ and $\langle A_{\rm V} \rangle  = 0.3$, and study the dependence on the turbulent parameters $\ms$ and $\fl$.
The effects of $I_{\rm UV}$, $\zeta$ and $A_{\rm V}$ are studied in \citetalias{Neufeld2016}. 
The LoS direction from the observer to the cloud (simulation) along which we integrate the column is aligned with the $z$ axis.
$z$ is also the axis along which the radiation propagates (i.e., as we assume slab geometry).
In the Appendix we explored other LoS orientations, along $x$ or $y$, and find that the results are weakly sensitive to the LoS direction.
%Thus, $x,y$ coordinates denote different LoS through the simulation boxes.

We post process each of the simulations as follows:
\begin{enumerate}
\item The 3D density field, $n(x,y,z)$, is used as an input to the chemical model.
\item 
For each LoS, $(x,y)$, we solve Eqs.~(\ref{eq: H_H2 steady state}-\ref{eq: diss rate})\footnote{This requires iterations since at any point $x({\rm H_2})$ depends on the entire H$_2$ profile on both sides (through the H$_2$ self-shielding function).
Clouds of sufficiently large column density may be modeled by a combination of two semi-infinite slabs, for which the H and H$_2$ profiles may be obtained analytically by integrating over the H$_2$ shielding function \citep[][]{Bialy2016a}.
However, this approximation is justified only if $N({\rm H}) \gg 2/\sigma_g \ln[D_0G/(2Rn)+1] \approx 5 \times 10^{20}$ cm$^{-2}$ 
where $G \approx 3\times 10^{-5}$ \citep{Sternberg2014}.
Our clouds do not satisfy this condition and thus we must solve for the two-sided irradiation.} on
a fine logarithmic grid, and  then interpolate the results back onto the simulation box.
\item
Given $x({\rm H_2})(x,y,z)$ and $n(x,y,z)$, we use the lookup tables to 
obtain the ArH$^+$, OH$^+$ and H$_2$O$^+$ abundances in all the cells, $x(i)(x,y,z)$.
\item
We also integrate the abundances
along the LoS direction, $z$, yielding the species column densities for all LoS, $N(i)(x,y)$.
\end{enumerate}
We now have the abundance and column density PDFs for the six cases:
$\ms=(0.5, 2, 4.5)$ for $\fl =(0.08, 0.8)$.

% The slabs span all $(x,y)$, and $z$ goes from $z_0$ to $z_0+\delta z$. We draw 10$^6$ random numbers for $x,y$ and $z_0$, and calculate the abundances for these 10$^6$ LoS. 

% The local abundances 
% in the turbulent cloud are thus a function of the ratios as inputs to the
% table based on their local values.
% We tested this method by substituting the turbulent density profile
% for the constant density
% in the diffuse cloud modeling and confirmed we obtained  similar distributions
% of molecular ions.

\section{Results}
\label{sec: results}

% \begin{figure}
% 				\label{fig: profiles}
% 	\centering
% 	\includegraphics[width=0.5\textwidth]{plots/fig3_profiles.png} 
% 	\caption{
% The abundance profiles (species density normalized to hydrogen nuclei density) of H$_2$, ArH$^+$, OH$^+$, and H$_2$O, through the $\ms=4.5$ $\fl=0.1$ simulation, for three randomly picked LoS (different colored curves).
% The grey strip the solution for a uniform density gas with the a volume density equal to the mean density.
% 		}
% \end{figure}

In this section we present the probability distribution functions (PDFs) of the abundances $x(i)$, and the columns, $N(i)$.

%We also compare the simulated PDFs with observations (\S \ref{sub: obs}).
%While the observations are most directly related to the integrated column densities (or column density ratios), we can gain understanding by first considering the density structure within the cloud, and the corresponding species abundance profiles (i.e., the species densities versus cloud depth).

%In Figure \ref{fig: sim_box}, the density fluctuations have a typical size of 1/10 of the simulation box, as parameterized by the decorrelation lengthscale, $f_{\rm fluc}=0.1$ (see discussion in \S \ref{sub: MHD}).
%In simulations with $f_{\rm fluc}=1.0$ (not shown), a density fluctuation has a spatial extent of order of the entire box. 
%As discussed in \S \ref{sub: MHD}, when considering this case, we produce a large number of realizations, spanning the entire density PDF.
%Thus, for both cases, $f_{\rm fluc}=0.1$ and 1.0, the density PDFs are similar, only the fluctuation sizes are different.

%The density fluctuations in the gas (as modeled through the MHD simulation), perturb the formation/destruction rates of the various species resulting in strong deviations from the uniform-density solution.

%\subsection{Abundance Profiles}
%\label{sub: Profiles}
\begin{figure*}[t]
	\centering
	\includegraphics[width=1\textwidth]{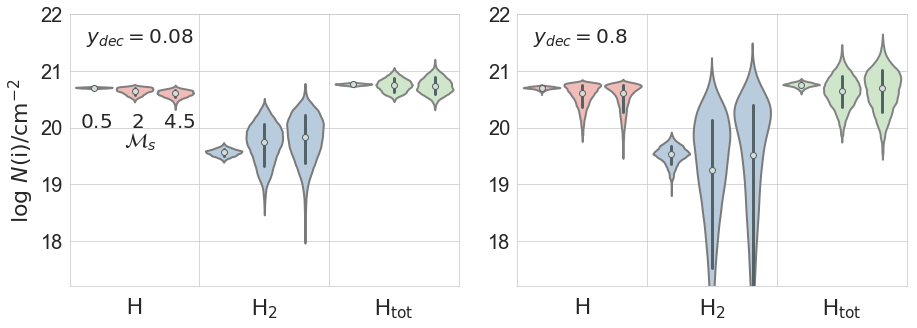} 
    \includegraphics[width=1\textwidth]{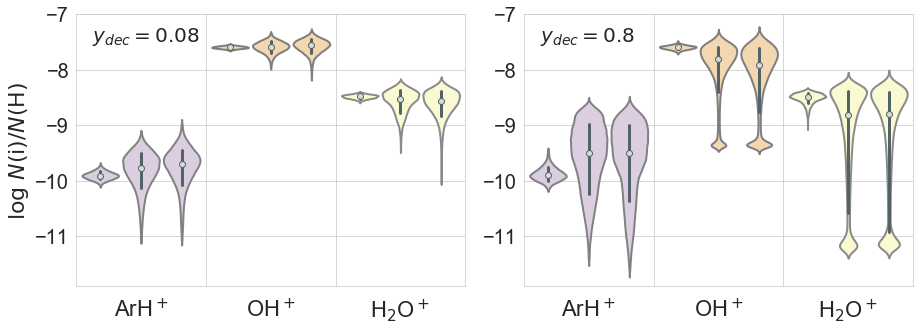} 
	\caption{
    Violin plots (where the width along the $x$ axis is $\propto$ probability density) of the H, H$_2$ and H$_{\rm tot}$ columns (top), and the ArH$^+$ OH$^+$ and H$_2$O$^+$ columns relative to the H column (bottom), for 
    simulations with $\fl=0.08$ (left) and $\fl=0.8$ (right), assuming $I_{\rm UV}=1$, $\zeta_{-16}=4$, $\langle n \rangle=30$ cm$^{-3}$, $\langle A_{\rm V} \rangle=0.3$.
    For each species, the triplet of violins corresponds to $\ms=0.5, 2, 4.5$, as indicated.
    For each violin, the white point, vertical line, and total vertical length, correspond to the  median, and the 68 and 99.7 percentile ranges about the median, respectively.  
		}
	\label{fig: pdfs one cloud}
\end{figure*}

\subsection{Abundance PDFs}
\label{sub: abundance PDFs}

To demonstrate the effect of the density fluctuations on the chemical structure, in Fig.~\ref{fig: abundance PDFs} we plot 
the joint-PDFs for $n$, $x({\rm ArH^+})$, $x({\rm OH^+})$ and $x({\rm H_2O^+})$ with $x({\rm H_2})$, and their marginal PDFs, for the subsonic $\ms=0.5$ (left) and the supersonic $4.5$ (right) simulations (both with $\fl=0.08$).
In the subsonic case the density PDF is narrow, with most cells having density $n\approx \langle n \rangle = 30$~cm$^{-3}$.
The H$_2$ distribution exhibits a spread, from $\log x({\rm H_2}) \sim -5$ to $-1$,  corresponding to regions near cloud boundaries and center.
The ArH$^+$, OH$^+$ and H$_2$O$^+$ PDFs as functions of $x({\rm H_2})$ follow the shapes of the constant density $n=30$ cm$^{-3}$ models (see Fig.~\ref{fig: x_vs_x2}), as they should.

In the supersonic case, strong density fluctuations are developed, and the $n$ PDF is wide.
The onset of H$_2$ self-shielding leads to a rapid growth in $x({\rm H_2})$ with increasing cloud depth (column density) resulting in a bimodal PDF.
At a given depth, higher gas densities result in more efficient H$_2$ formation and thus a positive correlation between $x({\rm H_2})$ and $n$.
At cloud boundaries, the correlation is linear, with $x({\rm H_2}) = Rn/(D_0/2)$.
This is evident as the thin diagonal density enhancement extending from $\log x({\rm H_2}), \log n = (-3, 3)$ to $(-6, 0)$.
For $n \lesssim 10$ cm$^{-3}$, there is an additional density enhancement, which also extends diagonally, but at an offset of factor of 2 to the left.
This corresponds to LoS that are so diffuse that radiation from the far side of the cloud penetrates, resulting in a factor of 2 higher dissociation rate.  

The density fluctuations in the $\ms=4.5$ simulation enhance the dynamic range of $x({\rm H_2})$ and thus the dynamic ranges of ArH$^+$, OH$^+$ and H$_2$O$^+$.
At a given $x({\rm H_2})$, the density fluctuations also lead to a spread in the ArH$^+$, OH$^+$ and H$_2$O$^+$ PDFs, through their dependencies on $\zeta/n$.
% The spread may be large or small as determined by the individual chemical formation/destruciton pathways of the species.
% For example, at small and large $\log x({\rm H_2})$ the ArH$^+$ PDF is very narrow 
% as the dynamic range in $n$ is small, and the 
% while at $x({\rm H_2})$ it spans
% and spans values within $\approx -10$ to -8 at $\log x({\rm H_2})=-3$, and finally falls to values -13 and below.
Since $x({\rm H_2})$ and $n$ are correlated, the PDF peaks as functions of $x({\rm H_2})$ no longer follow any of the constant density models.
For example, as $x({\rm H_2})$ increases from $10^{-4}$ to 0.5, the ArH$^+$ falls from $\approx 10^{-9}$ to $\lesssim 10^{-13}$, which correspond to the $n=30$ and 
300 cm$^{-3}$
models in Fig.~\ref{fig: x_vs_x2}, respectively.
  
The marginal PDFs of the molecular ions are composed of a broad and a narrow component.
In the case of OH$^+$ and H$_2$O$^+$ the narrow components correspond to formation on dust near cloud boundaries (where $x({\rm H_2})$ is small) which lead to characteristic OH$^+$ and H$_2$O$^+$ abundances, that depend weakly on $x({\rm H_2})$ or $n$.
In the case of ArH$^+$, the narrow PDF reflects the maximum in the $x({\rm ArH^+})$ versus $x({\rm H_2})$ relation (which occurs at $x({\rm ArH^+}) \sim 10^{-9}$ for $\langle n \rangle =30$ cm$^{-3}$).

\subsection{Column Density PDFs}
\label{sub: Column PDFs}

 In Fig.~\ref{fig: pdfs one cloud}
we show the column density  (or column density ratios) PDFs
% for ${\rm ArH^+, OH^+, H_2O^+}$ and 
% ${\rm H, H_2, H-tot}$, 
in the form of violin charts (where the width is proportional to the PDF dispersion), 
as obtained 
for all of our simulations, $\ms=0.5, 2$ and $4.5$, for $\fl=0.08$ (left) and 0.8 (right).

All PDF widths increase with increasing $\ms$
as a result of the density fluctuations.
The widths vary among species as is determined by the individual dependencies of $x(i)$ on $n$ (see \S \ref{sub: abundance PDFs}).
The column PDFs are further affected by the density as cells with higher density 
contribute more to the column, $N(i) = \int x(i) n \mathrm{d} z$.
% In addition, unlike abundance PDFs, for the column density PDF the density plays an extra role as a weighting, since $N(i) = \int x(i) n$.
For example, the $N({\rm H})$ PDFs are all very narrow because $x({\rm H})$ is anti-correlated with $n$, thus the cells where $x({\rm H})$ is high have a low contribution to $N({\rm H})$, and vice versa.
By contrast, $x({\rm H_2})$ correlates with $n$, and thus the $N({\rm H_2})$ PDFs are very wide.

% \subsubsection{The sonic Mach number} 
% \label{subsub: Ms
%  For example, the ArH$^+$ abundance is very sensitive to density fluctuations, as any increase in $n$  results in both (a) an increased H$_2$ abundance, and (b) a lowered effective ionization, $\zeta/n$, both of which act to decrease the ArH$^+$ abundance, and vice versa.
%  Thus the ArH$^+$ PDF is wide, with the LoS that have low densities producing the higher ArH$^+$ abundances, and vice versa. 
% Conversely, for H$_2$O$^+$, LoS with very low densities also produce low H$_2$O$^+$ abundances, as $x({\rm H_2O^+})$ becomes small at low $x({\rm H_2})$ (see Fig.~\ref{fig: abundance PDFs}).
% to the dependence of the abundances on density and H$_2$ content through the individual chemical network, 
% Importantly, the abundances, defined as $N(i)/N({\rm H})$, \footnote{We normalize to the atomic hydrogen and not the total gas column, as atomic hydrogen is an observable. However, in any case in our models, the total gas column is typically dominated by atomic hydrogen, with $N/N({\rm H})  \approx 0.75$.} are not affected by this effect.

The medians also show a dependence on $\ms$.
The median ArH$^+$ increases with $\ms$, both at $\fl=0.8$ and 0.08.
This is because the density field is characterized by a lognormal PDF in which low densities are more common than high (i.e., $n_{\rm med}<\langle n \rangle$).
These diffuse regions are deficient in H$_2$, and thus enhanced in ArH$^+$, leading to an overall increase in the median ArH$^+$ value.
This is evident by comparing the $\ms=0.5$ and 4.5 abundance PDFs in Fig.~\ref{fig: abundance PDFs}.
For $\fl=0.8$, the median OH$^+$ and H$_2$O$^+$ {\it decrease} with increasing $\ms$, as the H$_2$-poor gas 
{\it depresses} OH$^+$ and H$_2$O$^+$ formation.

At small $\fl$, the trends with $\ms$ are less pronounced as there are many density fluctuations along the sightlines, which, upon summation, partially cancel out (see \S \ref{subsub: fluc} below).
The sum is not evenly weighted as cells with
high density contribute more to the total integrated column. 
This is the reason why the median OH$^+$ trend is reversed at $\fl=0.08$:
While $x({\rm OH^+})$ is depressed in the low $x({\rm H_2})$ (low $n$) regions, the depression is sufficiently weak so that the $x({\rm OH^+})$-rich (high $n$) regions are able to (over)compensate through the larger weight they have in the integrated column.

%Sightlines with low densities also have low $x({\rm H_2})$ and are therefore enhanced in ArH$^+$ (recall, the ArH$^+$ abundance increases with decreasing $x({\rm H_2})$).
%On the other hand, high density sightlines sightlines with high density regions are (a) less common, and (b) even though the magnitude of their density fluctuations is more extreme, they barely affect $x({\rm ArH^+})$. This is because the $x({\rm ArH^+})$ dependence on $x({\rm H_2})$ is very weak at high H$_2$ abundances, $x({\rm H_2}) \gtrsim 0.1$.
%The final result is that the median ArH$^+$ increases with $\ms$.
%	

%For a lognormal $n$-PDF the median density follows
%\begin{equation}
%	x_{\rm med} \equiv \frac{n}{\langle n \rangle} =  \mathrm{e}^{\mu_{\ln x}} = \frac{1}{\sqrt{1 + (b \ms)^2 }} \ .
%\end{equation}
%The median density decreases with increasing Mach number, and therefore so does $x_{\rm H_2, med}$ (which correlates with the density).
%Finally, $x({\rm ArH^+})$ depends inversely on  $x({\rm H_2})$,and thus higher ArH$^+$ abundances are typically found in the high $\ms$ simulations.
%When calculating the column density of ArH$^+$, high density regions along the LoS (which thus have low ArH$^+$ abundances) will also contribute however,
%Similarly, high density regions lead to low ArH$^+$ abundances, but this effect does not cancel out with the former, because 

\subsubsection{The Role of $\fl$} 
\label{subsub: fluc}

The PDFs in Fig.~\ref{fig: pdfs one cloud}  depend strongly on the decorrelation scale, $\fl$ 
(see discussion in \S \ref{sub: MHD}).
Most prominent is the increase in the PDF dispersions as $\fl$ increases.
This is because when $\fl$ is large, the density fluctuations are coherent along the LoS (the number of fluctuations $\sim 1/\fl$).
Since the density still varies {\it between} LoS, there are large variations in the abundances among the sightlines and the PDFs are wide.
On the other hand, when $\fl$ is small, 
each LoS contains several ($\sim 10$) density fluctuations {\it along} it.
Upon integration, positive and negative fluctuations average-out, resulting in narrower PDFs.
The  H$_2$ PDF is particularly wide in the $\fl=0.8$ case, because 
at any point in the cloud, the H$_2$ is affected by all cells along the LoS through H$_2$ self-shielding and thus coherent density fluctuations strongly affect H$_2$.

Generally, because at large $\fl$ the density field is more coherent, the column PDFs more closely resemble the $x(i)$ PDFs, in this limit.
For example, in the $\fl=0.8$ case 
the OH$^+$ and H$_2$O$^+$ PDFs both have peaks at low values (at $\log N({\rm i})/N({\rm H}) \approx -9.3$ and $-11.1$), which are absent in the $\fl=0.08$ case.
These peaks correspond to low-density clouds with low $x({\rm H_2})$,
in which the OH$^+$ and H$_2$O$^+$ are formed through the dust-catalysis formation routes, reflecting the narrow peaks in the $x(i)$ PDFs in Fig.~\ref{fig: abundance PDFs}.
For $\fl=0.08$, these features disappear as the low-$x(i)$ regions are averaged over with  regions with higher $x(i)$, where the high-$x(i)$ regions have a larger weight.

%In this limit, the different LoS are more alike, as each LoS samples several random values from the $n$-PDF.

The dependence of PDF width on $\fl$ may be understood analytically for the case of the $N$ PDF (also in the case of $N({\rm H})$, see \citetalias{Bialy2017}).
Following \citetalias{Bialy2017}, 
we approximate the density field in two cells along the LoS as (a) two uncorrelated variables if $\Delta z>L_{\rm dec}$ where $\Delta z$ is the seperation along the LoS, or (b) assume full correlation if $\Delta z < L_{\rm dec}$ .
The number of fluctuations along a LoS is then
\begin{equation}
\mathcal{N} = 1+ 1/\fl \ .
\end{equation}
In this approximation, the column density is the sum of $\mathcal{N}$ independent variables, and its standard deviation obeys
\begin{equation}
\label{eq: sig_N}
\sigma_{N/\langle N \rangle} \simeq \frac{\sigma_{n/ \langle n \rangle}}{\sqrt{\mathcal{N}}} \simeq  \frac{b \ms}{\sqrt{1+ 1/\fl}} \ .
\end{equation}
The PDF dispersion increases with $\ms$ and with increasing fluctuation size,  $\fl$ (or with decreasing $\mathcal{N}$).
% For example, for $\fl =0.08$ and $\ms =(0.5, 2, 4.5)$, 
% Eq.~(\ref{eq: sig_N}) predicts
% $\sigma_{N/\langle N \rangle} = (0.045, 0.18, 0.41)$, respectively. 
% For $\fl =0.8$, the respective s.t.ds increases to $(0.11, 0.44, 1.00)$.
% This is in 10-30 \% agreement with the numerically calculated s.t.d.
 Eq.~(\ref{eq: sig_N}) is in 10-30\% agreement with our numerically calculated standard deviations for all six simulations.

\begin{figure*}[t]
	\centering
	\includegraphics[width=1\textwidth]{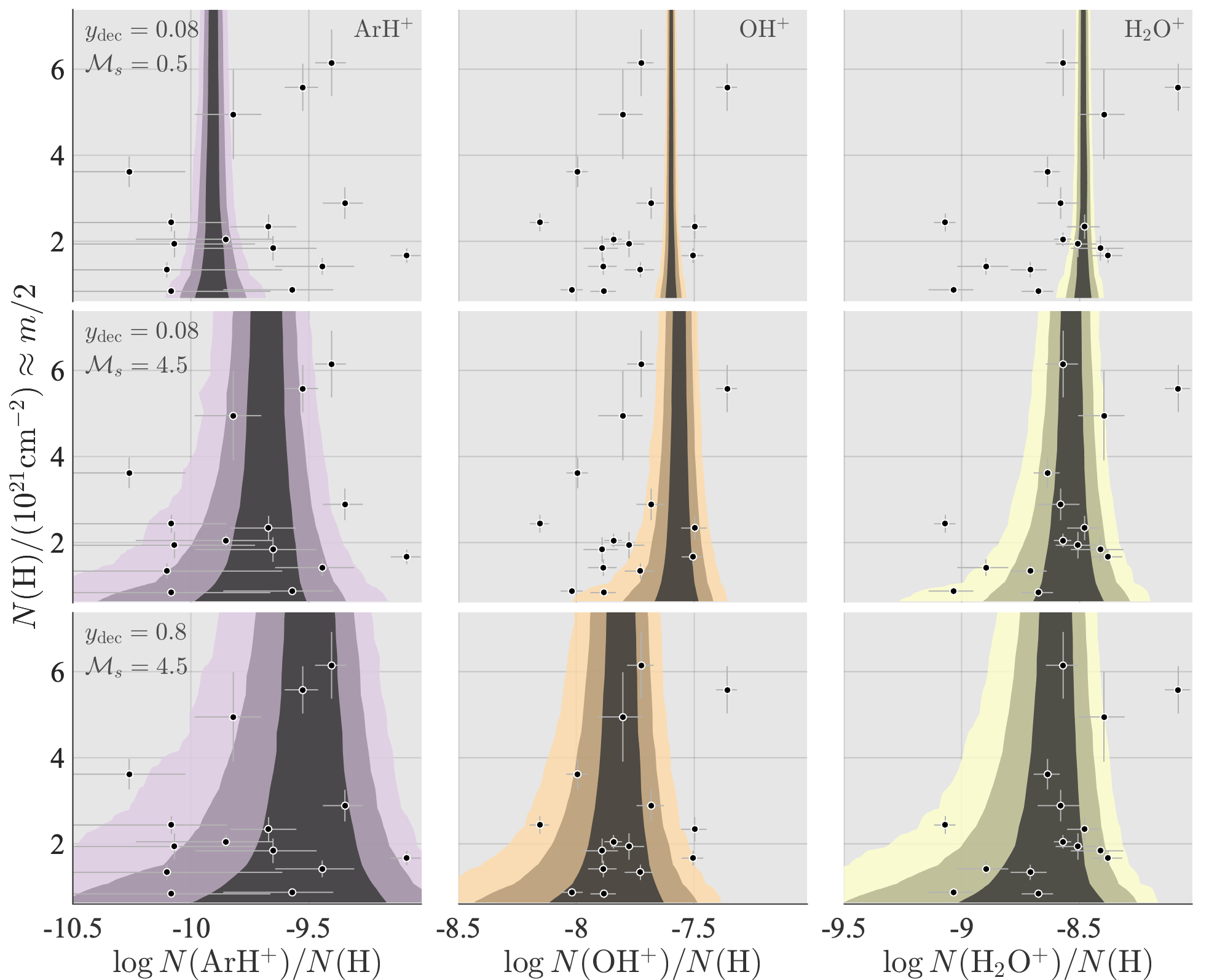} 
    \caption{
    The grand PDFs of ${\rm ArH^+, OH^+, H_2O^+}$ as functions of $N({\rm H})$ (which is $\propto$ to the number of clouds along the LoS, $m$.),
    for different ($\fl, \ms$) combinations.
    All models assume
    $I_{\rm UV}=1$, $\zeta_{-16}=4$, $\langle n \rangle=30$ cm$^{-3}$, $\langle A_{\rm V} \rangle=0.3$.
    In each panel, the three shaded regions correspond to the 68, 95, 99.7 percentiles about the median (at constant $N({\rm H})$).
    The observations are indicated by dots with errorbars.
		}
    \label{fig: pdfs many clouds}
\end{figure*}

In the limit $\fl \gg 1$,
$\sigma_{N/\langle N \rangle} \rightarrow \sigma_{n/\langle n \rangle}$ independent of $\fl$.
In this limit, each cloud corresponds to a single density fluctuation in the large-scale density field.
The column PDF of species may then be obtained directly from the uniform  
density models, weighting each model with the $n$-PDF.

% $\sigma_{X} = (0.04, 0.26, 0.34)$ 
% and $(0.09, 0.60, 0.91)$ for $\ms=(0.5, 2, 4.5)$, $\fl =0.08$ and $\fl=0.8$.

The decorrelation scale is related to the driving scale:
\begin{equation}
    \label{eq: Ldec}
    L_{\rm dec} =  \phi L_{\rm drive} \ ,
\end{equation}
or in dimensionless units, 
$\fl = \phi (L_{\rm drive}/L_{\rm cloud})$.
 In our simulations $\phi=0.2$
 (see \S 4.2 in \citetalias{Bialy2017}),  in agreement with analytic and numerical studies which find $\phi \approx 0.1-0.3$\footnote{
 For a line-width size relation with exponent $1/2$, the sonic scale $L_s = L_{\rm drive}/\ms^2$ \citep[][see \S 2.1.3]{McKee2007}, and is   typically $\ll L_{\rm dec}$. For example, for $\ms=4.5$, $L_s/L_{\rm drive}=0.05$, while $L_{\rm dec}/L_{\rm drive}=0.2$.
 } 
 \citep{VazquezSemadeni2001, Fischera2004, Kowal2007}.
  
The dependence of the species PDFs on $\fl$ may be potentially used as a method to constrain the turbulence driving scale in clouds (\citetalias{Bialy2017}).

%In the limit $\fl<1$, the PDFs are also affected by a weighting effect, as each LoS contains several density fluctuations, where the positive fluctuations have a larger weight as they have larger gas column density.
%For this reason 

% The increase in dispersion  with increasing fluctuation size is studied in detail in \citet{Bialy2017}, where it was shown that the distribution of $x_l$, the mean densities averaged along the LoS, increases with $\fl$. 
% The standard deviation obeys $\sigma_{xl} = \sigma_x \times (1+\fl^{-1})^{-1/2}$  (see their Eq.~22 \footnote{The notation in \citet{Bialy2017} is related to $\fl$ as follows: $L_{\rm dec}/L_{\rm cloud} \equiv \fl$.}), where the 1/square-root dependence follows from the 
%central limit theorem (see also, cite XXX).
%The standard deviation of the H column also increases with increasing $\fl$, but the analytic expression is more involved and also includes a dependence other parameters, such as $I_{\rm UV}/n$ and $\sigma_g$ through the chemistry of H and H$_2$ (see Eq.~38 in \citealt{Bialy2017}).
%Likewise, the dispersion in ArH$^+$, OH$^+$, and H$_2$O$^+$, increases with $\fl$ but also depends on the physical properties of the gas.

\section{Comparison to Observations}
\label{sec: obs}

% Here we compare the observed abundances of ArH$^+$, OH$^+$, and H$_2$O$^+$ (their column ratios, $\log_{10} [N(i)/N({\rm H})]$), to the theoretical PDFs.

\subsection{Observations}

The three molecular ions discussed here, ${\rm OH^+}$, ${\rm H_2O^+}$ and ${\rm ArH^+}$, have all been observed extensively in the diffuse Galactic ISM.   Absorption-line observations of their submillimeter rotational transitions near 972, 1115 and 618~GHz, respectively, have been carried out by the {\it{Herschel Space Observatory}} towards several regions of massive star formation that serve as background continuum sources.  Such observations typically reveal multiple absorption components arising in diffuse foreground gas that is unassociated and spatially-separated from the continuum sources; thanks to the differential rotation of the Galaxy, multiple diffuse clouds along each sight-line may be distinguished kinematically.  At the typical density in the diffuse ISM, these molecular ions are found primarily in the ground rotational state; thus, the observations yield robust estimates of the molecular column densities that do not depend on precise knowledge of the gas density or of the rate coefficients for collisional excitation.   Because ${\rm OH^+}$ and ${\rm H_2O^+}$ have rotational lines that show hyperfine structure, a deconvolution must be performed prior to the determination of their column densities \citep{Indriolo2015}. 

From {\it Herschel} observations reported \citep{Indriolo2015, Schilke2014} in previous literature, \citet[][hereafter \citetalias{Neufeld2017b}]{Neufeld2017b} identified fifteen velocity intervals for which all three molecular ions were detected within the spectra of 4 background continuum sources: W31C, G34.3, W49N, and W51e.  The choice of velocity intervals is somewhat arbitrary, in that the observed absorption cannot be reliably decomposed into the contributions from individual clouds; thus each velocity interval may contain multiple diffuse clouds.  The molecular column densities for these velocity intervals, along with ancillary estimates of the HI column densities derived from  21~cm observations \citep{Winkel2017}, comprise the observational data with which we compare the model predictions.

\subsection{The Grand PDFs}

Since the observations are probing through crowded regions in the Galactic plane, 
the sightlines likely contain several clouds superimposed along the LoS \citep[e.g., ][]{Bialy2017b}.
This is also supported by the observed spectra which show several distinct peaks, and is expected 
based on the large 
observed column densities $N({\rm H})_{\rm obs}=0.84-6.1 \times 10^{21}$ cm$^{-2}$. 
For our turbulent cloud models $\langle N({\rm H}) \rangle \approx 4 \times 10^{20}$ cm$^{-3}$
giving $m=2-15$ clouds along the LoS.

To compare our model with observations, we autoconvolve our single-cloud column PDFs (\S \ref{sub: Column PDFs}) as follows.
\begin{enumerate}
    \item For each simulation we randomly draw 10$^4$ LoS, and calculate the various column densities.
    \item We repeat step (1) $m$ times, and for each LoS we co-add the respective columns.
    This produces the ``$m$-cloud PDF", for $m$  superimposed clouds along the LoS.
    We do that for any $m \in [1,2,...,24]$\footnote{Here $m_{\rm max}=24$ is chosen to satisfy $m_{\rm max} \gg N({\rm H})_{\rm obs, max}/\langle N({\rm H}) \rangle \approx 15$.}.
    \item We stack all the $m$-cloud PDFs with equal weighting to produce 
    a single ``grand" PDF (for each species), which accounts for the superposition of several (unknown number) of clouds along the LoS.
    % As we apriori do not know the number of clouds\footnote{Because of the density fluctuations the total column varies among LoS in the simulation, and thus the number of clouds may also vary.}, we do not choose any particular value for $m$.
    % Instead, we assume that there is an equal probability for 
    % any value for $m$.
    % To this end we produce $m$-cloud PDFs (steps 1-2) for various $m$ values and sum the $m$-cloud PDFs with equal weighting.
\end{enumerate}

We show the grand PDFs as functions of $N({\rm H})$ in Fig.~\ref{fig: pdfs many clouds}
for (a) $\ms=0.5$, $\fl=0.08$, (b) $\ms=4.5$, $\fl=0.08$, and (c)
$\ms=4.5$, $\fl=0.8$.
In each panel, the three strips enclose the 68, 95 and 99.7 percentiles about the median (at constant $N({\rm H})$).
The observations are shown as dots with errorbars.
Here,
% with $N_{\rm H, 21}$ on one axis and $N(i)/N({\rm H})$ on the other, 
$N({\rm H})$ on the $y$-axis encodes information regarding the number-of-clouds, with $m \approx 2 N({\rm H})/(10^{21} \ {\rm cm^{-2}})$.
This is thanks to the fact that the $N({\rm H})$-PDFs of a single cloud are very narrow (see Fig.~\ref{fig: pdfs one cloud}).
On the other hand, from an observational point, $N({\rm H})$ is a direct observable.
On the $x$-axis, since the columns are normalized to $N({\rm H})$, the median values are independent of $m$ (or $N({\rm H})$), and all align up with the corresponding single-cloud values.

The dispersions of the PDFs (across the $x$ axis) decrease with increasing $m$, as $1/\sqrt{m}$, 
(see discussion in \S \ref{subsub: fluc}).
The PDF shapes approach Gaussians with increasing $m$, as required by the central limit theorem.
Importantly, Fig.~\ref{fig: pdfs many clouds} allows the comparison of observations with theory, even though the different observational sightlines may have different number of clouds, and thus a different predicted PDF width.

%with $\sigma_{N({\rm H})}/\langle N({\rm H}) \rangle < 2 \%$, and $\langle N({\rm H}) \rangle \approx N_{\rm H, med} \approx 5 \times 10^{20}$ cm$^{-2}$, 
Comparing the three cases in Fig.~\ref{fig: pdfs many clouds} we observe the familiar trends as for the single-cloud PDFs (\S \ref{sub: Column PDFs}):
(1) the increase in the PDF dispersions as the gas becomes supersonic and density fluctuations develop,
(2) the further increase in the dispersions as the decorrelation scale increases, $\fl$ and
(3) the shift of the ArH$^+$ median to higher values, and the OH$^+$ and H$_2$O$^+$ medians to lower values,  as $\ms$ and $\fl$ increase.

The model that explains best the observations is the $\ms=4.5$, $\fl=0.8$ model
% For example, in model (a) [6, 6, 3] out of the 15 observed  abundances (for, [ArH$^+$, OH$^+$, H$_2$O$^+$], respectively) are enclosed  within the 95 percentile contours of the model PDFs.
% For the $\ms=4.5$, $\fl=0.1$ model, the corresponding numbers increase to [10, 5, 9]/15.
% Finally, for $\ms=4.5$, $\fl=1.0$, [14, 11, 13] out of the 15 points are encLoSed within the  95 percentile contours.
in which the PDF dispersions are largest, and where the $N({\rm ArH^+})/N({\rm OH^+})$ ratios are highest.
The large decorelation scale combined with Eq.~(\ref{eq: Ldec}) suggests that the density fluctuations are driven on scales $L_{\rm drive} \gg L_{\rm cloud}$.
However, even in this model the observational scatter is too large requiring either still larger $\ms$ and $\fl$, or additional fluctuations in the non-turbulent parameters as we discuss below.

\section{Discussion}
\label{sec: Discussion}
%{\color{magenta}
%Here we discuss the limitations of our model, and compare our results to previous work
%}

\subsection{Variations in non-turbulent parameters}
\label{sub: var non  turb}
    As our focus in this paper is on the effect of the turbulence induced density perturbations, we fixed the ``non-turbulent parameters", $I_{\rm UV}$, $\zeta$, $\langle n \rangle$ and $\langle A_{\rm V} \rangle$, and studied the behavior as function of the ``turbulent parameters", $\ms$ and $\fl$.
    However, in the real ISM, the non-turbulent parameters may vary between clouds inducing additional variations to the abundances.
    In fact, even in our high-$\ms$ high-$\fl$ model, in which the dispersion is maximal, it is still not large enough to account for the scatter in the observations. 
    This suggests additional fluctuations in the non-turbulent parameters, and/or a higher Mach number and/or higher decorrelation lengths.

    It is important to keep in mind that the abundances do not depend on $I_{\rm UV}$, $\zeta$ and $\langle n \rangle$ separately, but instead are determined by the two ratios $I_{\rm UV}/\langle n \rangle$ and $\zeta/\langle n \rangle$.
    Therefore, if there exist positive correlations between the UV intensity, CRIR, and density, any fluctuation in each of these parameters will result in only a small fluctuation in their ratios.  
    Indeed, under thermal steady-state conditions, 
    the density of the cold neutral medium (CNM, $T \sim 100$ K)  is predicted to be positively correlated with $I_{\rm UV}$ \citep{Wolfire2003, Bialy2019}. 
    $\zeta$ and $I_{\rm UV}$ may also correlate given that both are produced by 
    massive stars (the latter in the supernovae remnants, at the stars' death), however, this  would depend on the CRs diffusion scale.
    
    Additional independent observational constraints of these parameters, their fluctuations, and correlations, are of great interest 
and will allow to reduce the degeneracy with the turbulent parameters.
    In the absence of additional measurements, the observed dispersion may still be used to place upper limits on the turbulent parameters and on the dispersion in the non-turbulent parameters.

        % Finally, our turbulent cloud model predicts that when  $\ms$ and $\fl$ are high, not only that the variance in species abundances becomes large, but also that the ArH$^+$ systematically increases while the OH$^+$ and H$_2$O$^+$ decrease, in agreement with observations.
    % The high observed ratios of ArH$^+$ to OH$^+$ and H$_2$O$^+$ were discussed previously by \citet{Neufeld2016}. 
    % They  considered a large set of (constant density) PDR models, with varying $A_{\rm V}$, $I_{\rm UV}/n$ and $\zeta/n$), and found that every model that reproduces the observed OH$^+$ and H$_2$O$^+$ abundances, always underpredicts the ArH$^+$, and vice versa.
    % This occurs because ArH$^+$ thrives in regions of where H$_2$ is very low (i.e., low density or low $A_{\rm V}$), while OH$^+$ and H$_2$O$^+$ peak at H$_2$ fractions, $\sim 0.1$.
    % To alleviate this discrepancy, \citet{Neufeld2016} proposed a two-cloud model of low and high $A_{\rm V}$ clouds.
    % In our turbulent models the ArH$^+$ abudnance and 

\subsection{Clouds and Density Fluctuations}   
\label{sub: cloud}
    What defines ``a cloud" in a turbulent ISM?
    Generally, one may define a cloud as a region that exceeds some threshold density, or mass.
    However, with this definition the cloud characteristics: its mass, size and internal structure, would depends on the (somewhat arbitrary) density/mass threshold value.
    Instead, in this study we defined the cloud based on the UV (more particularly, the LW) radiation field.
    We explicitly assumed that each cloud is irradiated by the mean  radiation field, 
    such that each cloud self-shields itself and does not affect the other clouds.
    For this condition to be fulfilled, the clouds need to be sufficiently separated in space (or in velocity space).
    Such a description of the ISM, as an ensemble of externally irradiated clouds, is useful but 
    is clearly an over-simplification of the real ISM.
    
    It is also worth reiterating the importance of taking into account the superposition of clouds along the LoS   (\S \ref{sec: obs}).
    While the mean and median abundances relative to H are the same for a single cloud and several superimposed clouds, the dispersion {\it is} sensitive to the number of clouds, and it decreases as $\simeq 1/\sqrt{m}$.
    While for $m=1$ the shapes of the PDFs are highly irregular, we find that as $m$ increases the PDFs approach Gaussians, as predicted by the central limit theorem.
    
\subsection{Thermal Phases of the ISM} 
\label{sub: phases}  
In this study we used isothermal MHD simulations, which mimic a single phase medium (the CNM).
In the real ISM, cooling/heating processes introduce a thermal instability which results in a 
multiphasic medium:
 the warm and cold neutral media (WNM, CNM), $T \approx (6000,60)$ K, $\langle n \rangle \approx (0.3,30)$ cm$^{-3}$ \citep[][]{Field1969, Wolfire1995, Wolfire2003, Bialy2019}, where intermediate temperatures are thermally unstable  \citep{Field1965}.
MHD simulations which include heating-cooling processes find a bimodal PDFs of  $T$ and $n$, with a non-negligible mass within the unstable region \citep[e.g.,][]{Piontek2007, Audit2010, Walch2011a, Saury2014}, where the bimodality is less pronounced in strongly turbulent gas
\citep{Gazol2013}.

How a bimodal PDF may affect the chemistry?
This critically depends on the mixing-scale of CNM/WNM structures.
Let $\lambda_{\rm c}$ be a characteristic  CNM scale, and consider two limiting cases:
(a) $\lambda_{\rm c} \ll L_{\rm cloud}$, and
(b) $\lambda_{\rm c} \gtrsim L_{\rm cloud}$,
where $L_{\rm cloud}$ is defined so that it is externally irradiated by LW radiation (see \S \ref{sub: cloud}).
In case (a), the CNM and WNM are mixed on small scales, and each cloud contains both phases, i.e., the entire bi-modal PDF is sampled in each cloud.
In this case, there would be more  H$_2$-poor gas compared to the pure CNM case, 
which in turn would affect the abundances of other species.
On the other hand, for case (b), the scale is large, so that each cloud samples only a part of the bimodal PDF, e.g., just the higher CNM densities. 
Our study thus resembles case (b). 
This discussion highlights the importance 
of determining the scales of WNM-CNM structures in the ISM
\citep[e.g., ][]{Heiles2003, Choi2012, McCourt2018, Waters2019}.

Even if an individual cloud is described by only the CNM phase,
if $L_{\rm drive} \gg L_{\rm cloud}$,   the driving
may take place in a multiphase CNM-WNM medium.
Both the $L_{\rm drive}-L_{\rm dec}$ and the $\sigma_{n/\langle n \rangle}-\ms$ relations  are derived from idealized isothermal simulations.
In future studies, it is important to generalize these relations to the more realistic case of driving in a multiphase medium.

\subsection{The H$_2$ self-shielding function}   
\label{sub: H$_2$ self-shielding}
The H$_2$ self-shielding function of \citet{Draine1996} and \citet{Federman1979} depend on the 
Doppler broadening parameter, $b$.
The assumption is that the gas is turbulent on small scales (compared to the length-scale over which the H$_2$ abundance change), aka, micro-turbulence.
Micro turbulence broadens the H$_2$ lines, in a similar manner to thermal broadening.
In appendix \ref{appendix doppler} we study the dependence on $b$ and show that this dependence is weak.

However, turbulence induces gas motions that may be correlated on various scales, up to the driving scale, and the micro-turbulence assumption may fail.
 \citet{Gnedin2014} have suggested an alternative shielding function that approximates this effect, however it still lacks the information on the velocity correlations, and the density fluctuations.
 On the other hand direct computations of H$_2$ shielding are possible given the velocities and densities in the simulation cells, however, this is computationally expansive as it requires radiative transfer calculations with high spectral resolution.
The effects of a turbulent velocity field on the H$_2$ column are expected to be important for clouds of small total column (for $I_{\rm UV}=1$ and $n=3-300$ cm$^{-3}$, $N \sim 10^{18}-10^{20}$ cm$^{-2}$, see Fig.~8 in
 \citet{Bialy2016a}, and for large velocity dispersions, with gas motions that are coherent over large scales.
 We plan to investigate the question of H$_2$ line absorption in turbulent gas elsewhere, with the goal of deriving an all-purpose self-shielding function.

\subsection{Comparison to Previous Work}
\label{sub: previous work}
In a previous study, \citetalias{Bialy2017} studied the atomic-to-molecular transition and the HI column PDF. 
They presented numerical results and  provided analytic formula for the $N({\rm H})$ dispersion as a function of \fl and $\ms$ (and the effective dissociation parameter, see their Eq.~38).
For similar reasons to those discussed here (\S \ref{subsub: fluc}), 
the dispersion in $N({\rm H})$ increases with $\ms$ and $\fl$.
 \citetalias{Bialy2017} applied their model  
to 21 cm observations of HI columns towards the Perseus molecular.
As the $N({\rm H})$ values towards Perseus are relatively uniform \citep[][see also \citealt{Imara2016} for more Galactic clouds with narrow HI PDFs]{Lee2012,  Bialy2015c}, \citetalias{Bialy2017} concluded that  $\fl$ is small.
Conversely, in this study we find that the dispersions  in the abundances of ArH$^+$, OH$^+$, and H$_2$O$^+$, over the various LoS are relatively large, requiring a large $\fl$.

The large dispersion observed here which is not seen in Perseus, may suggest that non-turbulent parameters, i.e., $\zeta$ or $I_{\rm UV}$, contribute considerably to the observed scatter in the  abundances.
However, a direct comparison  is complicated as the 
observations in the two studies probe different kind of clouds, and in different Galactic environments.
Furthermore, other factors also affect the dispersion, such as the number of clouds along the LoS, and the multiphase structure of the ISM (see \S 7-8 in  \citetalias{Bialy2017} for a further discussion). 

Our adopted value, $\zeta_{-16} = 4$, optimizes the fit of our supersonic high $\fl$ model to the measured column densities of OH$^+$, H$_2$O$^+$, and ArH$^+$.
Our value is a factor of $(1.4-2)$ higher than that derived previously by \citetalias{Neufeld2017b}.
For the subset of 15 sources considered here (in  which ArH$^+$ was detected), and using their two-cloud population model with uniform densities, they derived $\langle \log \zeta/n_{50} \rangle = -15.25$ with a standard error on the mean $\pm 0.07$ (see their Table 2).
For $n=30$ cm$^{-3}$ this corresponds to $\langle \zeta_{-16} \rangle \approx (2.0-2.8)$,
where we included a downward correction to  account for the higher $I_{\rm UV}/n$ assumed here, following the relation $\zeta \propto (I_{\rm UV}/n)^{0.7}$ (see \S 3.4 in \citetalias{Neufeld2017b}).
For a further comparison to \citet{Indriolo2012}, see Table 2 in \citetalias{Neufeld2017b}.

\section{Conclusions}
\label{sec: conclusions}
We studied how turbulence-induced density fluctuations affect the chemical structure of diffuse clouds, our conclusions are:
\begin{enumerate}
\item ArH$^+$, OH$^+$, and H$_2$O$^+$ are sensitive probes of the 
H$_2$ abundance, which is in turn very sensitive to density fluctuations, because of H$_2$ self-shielding. 
\item When $\ms < 1$, the density fluctuations are weak, and the resulting abundances converge to those predicted by the former uniform-density models.
For supersonic gas, the density fluctuations become strong, and as a result the H$_2$, ArH$^+$, OH$^+$ and H$_2$O$^+$ abundances in the cloud span a large range.
\item The shapes of the $x(i)$-PDFs are irregular, typically double-peaked, and are determined through the unique chemical formation pathways, e.g., OH$^+$ and H$_2$O$^+$ gas-phase formation versus formation on dust.
\item The column density PDFs become wider as $\ms$ increases, and also as  $\fl$ increases.
At high $\fl$, the density field is more coherent, enhancing the effect of the density fluctuations on the chemical structure.
On the other hand, when $\fl$ is small, positive and negative fluctuations partially average out.
\item The medians are also affected by the density fluctuations.
With increasing $\ms$ and $\fl$, the median ArH$^+$ increases, while OH$^+$ and H$_2$O$^+$ decrease.
When $\ms$ and $\fl$ are high, regions with low H$_2$ abundance are prevalent, which favor ArH$^+$ over OH$^+$ and H$_2$O$^+$.
\item The observed abundances have a considerable scatter, and high ArH$^+$-to-OH$^+$ and ArH$^+$-to-H$_2$O$^+$ ratios.
These suggest supersonic gas with large decorrelation-scales, which in turn correspond to large-driving scale, $L_{\rm drive} \gg L_{\rm cloud}$.
\item The abundances also depend on $\zeta/\langle n \rangle$, $I_{\rm UV}/\langle n \rangle$ and $A_{\rm V}$, and the observed scatter may result from fluctuations in these parameters.

\end{enumerate}
More independent observations are needed to break the degeneracy between the turbulent and non-turbulent parameters. On the theoretical part, the next steps are to (a) develop more realistic models that include the interplay of turbulence and thermal phases, and (b) a more accurate H$_2$ shielding function that applies to turbulent medium.
With these advances, observations of molecular ions may be used as a robust tool to constrain the  driving and strength of interstellar turbulence, and/or variations in the interstellar fields: the UV intensity and the cosmic-ray ionization rate.
  
% \appendix

% \section*{A. Lookup-Tables and Analytic Fit}

% We obtain the abundances in the following way.

% \begin{figure}[t]
% 	\centering
% 	\includegraphics[width=0.5\textwidth]{fig1_HI_profiles_hom.png} 
% 	\caption{
% caption
% 		}
% 		\label{fig: sim_box}
% \end{figure}

\acknowledgements
We thank the anonymous referee for valuable comments and suggestions that improved this work.
We thank the Center for Computational Astrophysics at the Flatiron Institute for hospitality and funding where some of this research was carried out.
This work was also supported by the German Science Foundation via DFG/DIP grant STE 1869/2-1 GE 625/17-1 at Tel-Aviv University.

\appendix
\section{A.~Dependence on the Doppler broadening parameter}
\label{appendix doppler}
\begin{figure}[h]
	\centering
	\includegraphics[width=1\textwidth]{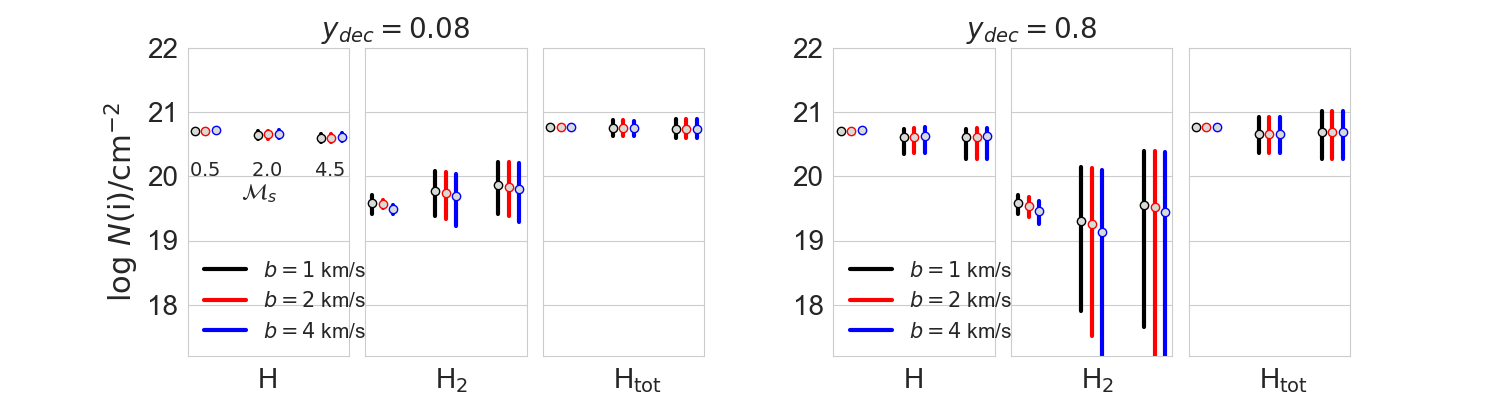} 
	\includegraphics[width=1\textwidth]{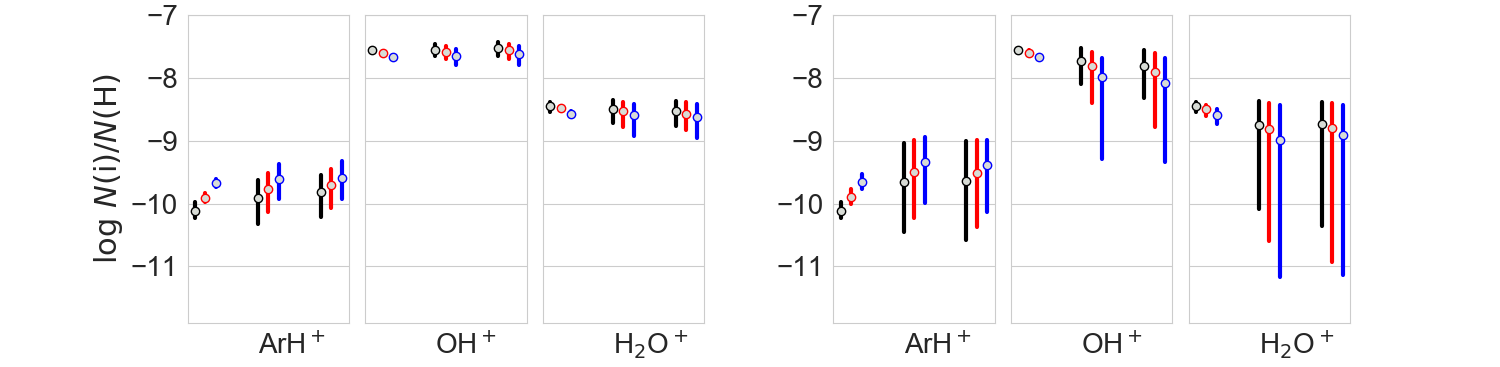} 
	\caption{
The median (points) and the 68 percentile range (bars) of the column density abundance PDFs, for our entire simulation set, and assuming our fiducial $I_{\rm UV}$, $\zeta$ and $\langle n \rangle$ and $\langle A_{V} \rangle$.
For each simulation, the three bars correspond to different  Doppler broadenings, $b=1$, 2, 4 km/s.
		}
    \label{fig: effect_of_b}
\end{figure}

The Doppler broadening parameter, $b$, enters the H$_2$ self-shielding function.
In Fig.~\ref{fig: effect_of_b} we show the PDF median and dispersion for all the species and for all the simulations considered here, for different $b$ values.
As $b$ increases, the difference in velocities of gas elements along the LoS increases and self-shielding becomes less efficient.
As a result, the H$_2$ column decreases.
However, this dependence is very weak because absorption by the H$_2$ line Doppler cores is effective at relatively small H$_2$ columns $N({\rm H_2}) \approx 10^{16}-10^{17}$ cm$^{-2}$, corresponding to total columns $N \sim 10^{18}-10^{20}$ cm$^{-2}$ (for $I_{\rm UV}=1$ and $n=3-300$ cm$^{-3}$; see Fig.~8 in
 \citet{Bialy2016a}.
Since the clouds considered here have $\langle N \rangle = 5.8 \times 10^{20}$ cm$^{-2}$, absorption by the Doppler cores is sub-dominant.
Instead, H$_2$ self shielding is dominated by the Damping wings (i.e., Lorentz broadening) independent of $b$.
For similar reasons, the H and the total columns are insensitive to $b$, in agreement with the finding of \citetalias{Bialy2017}.
 
The molecular ions, show a stronger dependence on $b$, especially, ArH$^+$.
 This is because the ions, and especially ArH$^+$, 
 are formed efficiently in regions of low $x({\rm H_2})$.
 These regions typically lie closer to the cloud edges where 
Doppler broadening is important. 
In any case, variations in $\ms$ and $\fl$ have a stronger effect on the PDFs.

\section{B.~Dependence on the LoS orientation}
\label{appendix LoS}
\begin{figure}[h]
	\centering
	\includegraphics[width=1\textwidth]{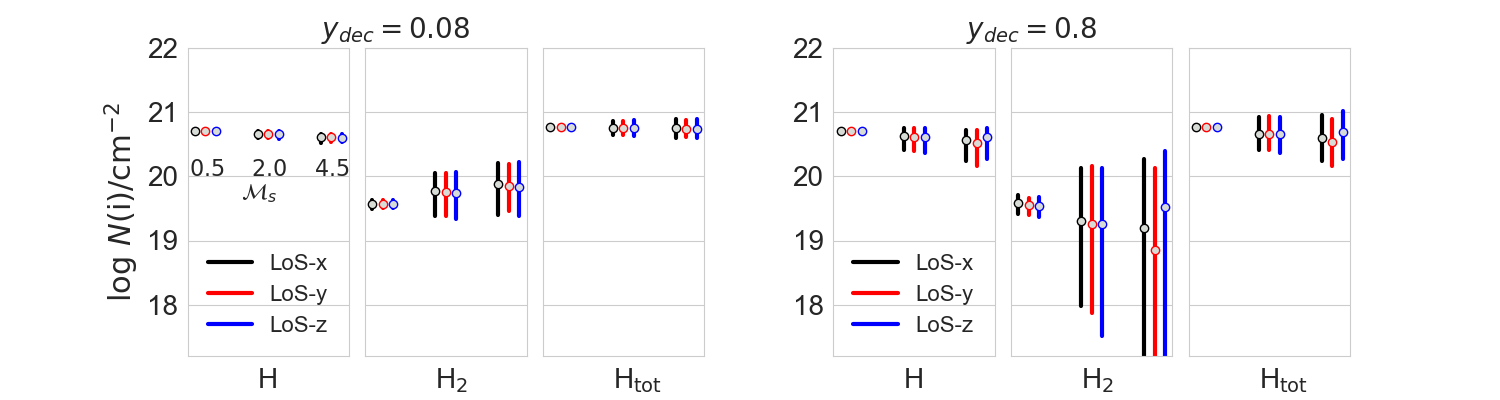} 
	\includegraphics[width=1\textwidth]{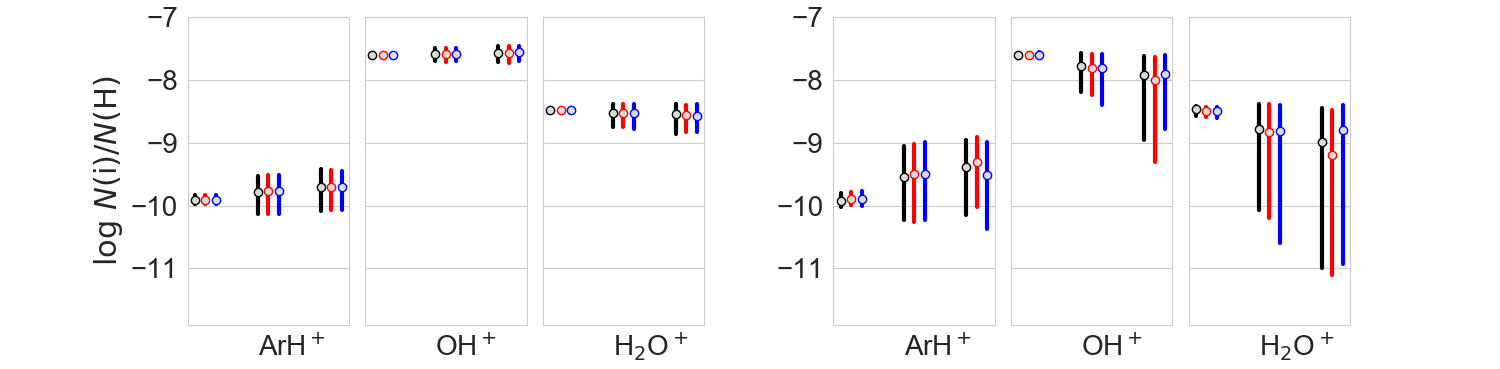} 
	\caption{
The median (points) and the 68 percentile range (bars) of the column density abundance PDFs, for our entire simulation set, and assuming our fiducial $I_{\rm UV}$, $\zeta$ and $\langle n \rangle$ and $\langle A_{V} \rangle$, and $b=2$ km/s.
For each simulation, the three bars correspond to different assumed orientations of the LoS: along $x$ (black), $y$ (red), and $z$ (blue). 
The $B$ field in the simulations is initialized along the $y$ axis. 
		}
    \label{fig: effect_of_los}
\end{figure}

The magnetic field in the simulations introduces a preferred direction (the  $B$ field is initiated along the $y$ axis).
In Fig.~\ref{fig: effect_of_los} we plot the PDF median and dispersion for all the species and for all the simulations considered here, for LoS parallel to $x$, $y$ or $z$.
As evident from the figure, the results are weakly sensitive to the direction of the LoS.

\end{document}